\documentclass{emulateapj}  % For more compact output
\usepackage{subfigure,verbatim}

\newcommand{\msc}[1]{\nobreak{\mbox{\scriptsize{\textsc{#1}}}}} % Ditto, but for small caps
\newcommand{\ud}{{\rm d}} % For upright differential

\newcommand{\grad}{\nabla}
\newcommand{\cross}{\times}
\newcommand{\bmath}[1]{\mbox{\boldmath{$#1$}}}
\def\deg{^\circ}
\def\comp{\,c/\omega_{\rm p}}
\def\ompt{\omega_{\rm p}t}
\def\thetacrit{\theta_{\rm crit}}
\newcommand{\gb}[1]{\gamma\beta_{\rm {#1}}}
\newcommand{\pan}[1]{\textit{\,#1}}
\newcommand{\tit}[1]{\textit{#1}}
\newcommand{\eq}[1]{eq.~(\ref{eq:#1})}
\newcommand{\fig}[1]{\ref{fig:#1}}
\begin{document}
\title{Particle Acceleration in Relativistic Magnetized Collisionless Pair Shocks: Dependence of Shock Acceleration on Magnetic obliquity}
\author{Lorenzo Sironi and Anatoly Spitkovsky}
\affil{Department of Astrophysical Sciences, Princeton University, Princeton, NJ 08544-1001}
\email{lsironi@astro.princeton.edu;\\ anatoly@astro.princeton.edu}

\begin{abstract}
We investigate shock structure and particle acceleration in relativistic magnetized collisionless pair shocks by means of 2.5D and 3D particle-in-cell simulations. We explore a range of inclination angles between the pre-shock magnetic field and the shock normal. We find that only magnetic inclinations corresponding to ``subluminal'' shocks, where relativistic particles following the magnetic field can escape ahead of the shock, lead to particle acceleration. The downstream spectrum in such shocks consists of a relativistic Maxwellian and a high-energy power-law tail with exponential cutoff. For increasing magnetic inclination in the subluminal range, the high-energy tail accounts for an increasing fraction of particles (from $\sim1\%$ to $\sim2\%$) and energy (from $\sim4\%$  to $\sim12\%$). The spectral index of the power law increases with angle from $-2.8\pm0.1$ to $-2.3\pm0.1$. For nearly parallel shocks, particle energization mostly proceeds via the Diffusive Shock Acceleration process; the upstream scattering is provided by oblique waves which are generated by the high-energy particles that escape upstream. For larger subluminal inclinations, Shock-Drift Acceleration is the main acceleration mechanism, and the upstream oblique waves regulate injection into the acceleration process. For ``superluminal'' shocks, self-generated shock turbulence is  not strong enough to overcome the kinematic constraints, and the downstream particle spectrum does not show any significant suprathermal tail. As seen from the upstream frame, efficient acceleration in relativistic (Lorentz factor $\gamma_0\gtrsim5$) magnetized ($\sigma\gtrsim0.03$) flows exists only for a very small range of magnetic inclination angles ($\lesssim34\deg/\gamma_0$), so relativistic astrophysical pair shocks have to be either nearly parallel or weakly magnetized to generate nonthermal particles. These findings place constraints on the models of AGN jets, Pulsar Wind Nebulae and Gamma Ray Bursts that invoke particle acceleration in relativistic magnetized shocks.
\end{abstract}

\section{Introduction}\label{sec:intro}
Collisionless shocks have been implicated in most nonthermal phenomena in the Universe. Models of nonthermal emission from Pulsar Wind Nebulae (PWNe), jets from Active Galactic Nuclei (AGN), Gamma-Ray Bursts (GRBs) and supernova remnants (SNRs) require a population of high-energy particles with power-law spectra, and collisionless shocks are the primary candidates to produce scale-free distributions of energetic particles. However, despite decades of research, particle acceleration in astrophysical shocks is still not understood from first principles. 

Shock acceleration of charged particles is thought to occur by means of two main mechanisms: Diffusive Shock Acceleration (DSA, or first-order Fermi acceleration) and Shock-Drift Acceleration (SDA).\footnote{In electron-ion magnetized shocks, Shock-Surfing Acceleration is also important, a mechanism by which ions repelled by the shock electric potential are accelerated by the background motional electric field \citep[e.g.,][]{lee_96}. In this work, we consider only electron-positron shocks, where the shock-surfing mechanism does not operate.}
In first-order Fermi acceleration, particles stochastically diffuse back and forth across the shock front and gain energy by scattering off magnetic turbulence embedded in the converging flows \citep[e.g.,][]{blandford_ostriker_78, bell_78,drury_83,blandford_eichler_87}.  SDA is the process whereby particles gain energy from the background motional  electric field $\mathbf{E}=-\mbox{\boldmath{$\beta$}}\times\mathbf{B}$ when gyrating across the shock front, while they drift parallel (or anti-parallel, according to their charge) to the electric field \citep[e.g.,][]{chen_75,webb_83,begelman_kirk_90}. 
SDA is a very fast process compared to DSA \citep{jokipii_82}; however, it is not very efficient without a means for the particles to remain close to the shock and encounter it more than once, thereby being repeatedly accelerated. It is generally accepted that when a sufficient amount of magnetic field turbulence exists on both sides of the shock, both mechanisms operate in tandem making the overall acceleration process very efficient. 

The efficiency of particle acceleration, namely the fraction of post-shock particles and energy stored in a suprathermal tail of the particle spectrum, may depend on the composition of the upstream flow (electron-positron, electron-ion or pairs-ion plasma), as well as its bulk Lorentz factor and magnetization. If the upstream medium is magnetized, an additional parameter is the obliquity angle that the upstream magnetic field makes with the shock direction of propagation. For strongly magnetized shocks, cross-field scattering is suppressed and the particle gyro-centers are constrained to slide along the field lines, which are advected downstream from the shock. It is then natural to define a critical magnetic obliquity above which particles sliding along the magnetic field should be moving faster than the speed of light in order to return upstream \citep{begelman_kirk_90}. Since efficient acceleration requires that particles repeatedly cross the shock, such ``superluminal'' geometries should be very inefficient for particle acceleration \citep{kirk_heavens_89, ballard_heavens_91}, unless there is strong scattering across field lines or appreciable fluctuations of the magnetic field at the shock, that may create local ``subluminal'' configurations. If the upstream medium is not turbulent by itself, magnetic field fluctuations may be generated by the high-energy particles accelerated at the shock that escape upstream. In turn, such magnetic waves could scatter the particles and mediate their acceleration. The highly nonlinear problem of wave generation and particle acceleration should then be addressed self-consistently.

Most progress in studying particle acceleration in magnetized shocks has been made using semi-analytic kinetic theory methods \citep[e.g.,][]{kirk_heavens_89, ballard_heavens_91,kirk_00, achterberg_01,keshet_waxman_05} or Monte Carlo test-particle simulations \citep[e.g.,][]{ostrowski_bednarz_98, niemiec_ostrowski_04, ellison_double_04}.
Both approaches study particle acceleration with simplifying (and somewhat arbitrary) assumptions about the nature of magnetic turbulence near the shock and the details of wave-particle interactions. 
In particular, it is found that superluminal shocks might accelerate as efficiently as subluminal configurations, but the required level of magnetic turbulence must be unreasonably large \citep{ostrowski_bednarz_02}. The open question is then if superluminal shocks can self-consistently produce such a high degree of turbulence.

Fully kinetic particle-in-cell (PIC) simulations provide a powerful tool for exploration of the fluid and kinetic structure of collisionless shocks on all scales, thus determining the nature of the shock-generated electromagnetic turbulence simultaneously with the spectrum of accelerated particles. PIC simulations of relativistic \tit{unmagnetized} shocks have been presented by \citet{spitkovsky_05, spitkovsky_08}, \citet{chang_08} and \citet{keshet_08}; \citet{spitkovsky_08b} has shown that unmagnetized shocks naturally produce accelerated particles as part of the shock evolution. One-dimensional PIC simulations  of \tit{magnetized} perpendicular pair shocks, i.e., with the magnetic field along the shock surface, have shown negligible particle acceleration \citep{langdon_88,gallant_92}, unless a sizable fraction of the upstream bulk kinetic energy is carried by heavy ions \citep{hoshino_92, amato_arons_06}. 

In this work, we investigate via first-principles two- and three-dimensional PIC simulations the acceleration properties of relativistic \tit{magnetized} pair shocks. We fix the upstream magnetic field strength and explore how the efficiency and mechanism for particle acceleration vary with magnetic obliquity. We find that subluminal configurations produce a population of energetic particles, and the downstream particle spectrum develops a high-energy power-law tail. We confirm that particle acceleration in superluminal shocks is extremely inefficient, meaning that self-generated turbulence is not strong enough to allow for significant cross-field diffusion of the particles. For subluminal shocks, we find that the acceleration efficiency increases as the magnetic obliquity approaches the boundary between subluminal and superluminal configurations. Also, by studying the trajectory of high-energy particles extracted from our simulations, we show that at low magnetic obliquities, where the motional electric field is still small, the acceleration process is mostly controlled by DSA, whereas particles are mainly accelerated via SDA when the magnetic field inclination is close to the superluminality threshold.

This work is organized as follows: in \S\ref{sec:setup} we discuss the setup of our simulations and the magnetic field geometry; in \S\ref{sec:struct} the shock structure and internal physics is investigated for some representative magnetic obliquities. The main results of our work, concerning the efficiency of shock acceleration as a function of magnetic obliquity, are presented in \S \ref{sec:spectra}. In \S\ref{sec:mechanism} we comment on the relative importance of different acceleration mechanisms (namely, DSA and SDA) as a function of the magnetic inclination angle. We summarize our findings in \S\ref{sec:disc} and comment on the application of our results to astrophysical scenarios.

\section{Simulation setup}\label{sec:setup}
We perform numerical simulations using the three-dimensional (3D) electromagnetic particle-in-cell (PIC) code TRISTAN-MP \citep{spitkovsky_05}, which is a parallel version of the publicly available code TRISTAN \citep{buneman_93} that was optimized for studying collisionless shocks. 

The shock is set up by reflecting a stream of pair plasma (the ``upstream'' plasma) off a conducting wall. The interaction between the injected and reflected beams triggers the formation of a shock, which moves away from the wall. This setup is equivalent to the head-on collision of two identical relativistic pair streams, which would form a forward and reverse shock and a contact discontinuity. In our simulations we follow one of these shocks, and replace the contact discontinuity with the conducting wall. The simulation is performed in the ``wall'' frame, where the ``downstream'' plasma behind the shock has zero bulk velocity along the direction of the upstream flow. We stress that the wall frame of our simulations does not in general coincide with the downstream fluid frame, since for oblique magnetic configurations the downstream plasma may have a velocity component transverse to the upstream flow, as we show in \S \ref{sec:struct}. 

We perform simulations in both two- and three-dimensional computational domains, and we find that most of the shock and acceleration physics is captured extremely well by two-dimensional simulations (see \S\ref{sec:disc} and Appendix \ref{append1} for a discussion). Therefore, to follow the shock evolution for longer times with fixed computational resources, we mainly utilize two-dimensional runs. All three components of particle velocities and electromagnetic fields are tracked, however. 

\begin{figure}
\begin{center}
\includegraphics[width=0.5\textwidth]{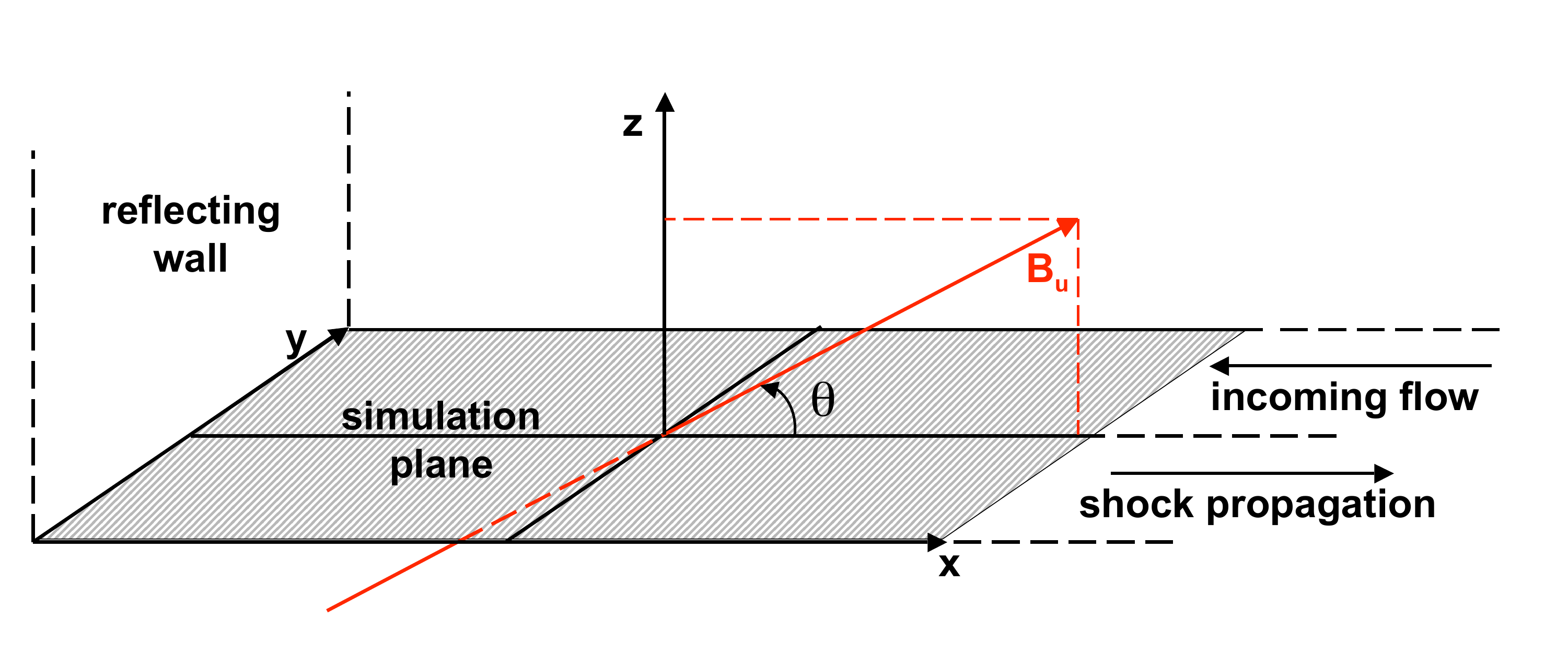}
\caption{Simulation geometry and configuration of the upstream magnetic field.}
\label{fig:simplane}
\nonumber
\end{center}
\nonumber
\end{figure}

We use a rectangular simulation box in the $xy$ plane, with periodic boundary conditions in the $y$ direction (Fig.~\fig{simplane}).  The box is 1024 cells wide (along $y$), corresponding to approximately $100\, c/\omega_{\rm p}$, where the plasma skin depth $c/\omega_{\rm p}=10$ cells. We also performed limited experiments with larger boxes, up to 4096 cells wide, obtaining similar results. In the expression above, $c$ is the speed of light and $\omega_{\rm p}\equiv\sqrt{4\pi e^2 n_{\rm u} /(\gamma_0 m)}$ is the relativistic plasma frequency of the upstream flow; $n_{\rm u}$ is the upstream number density measured in the wall frame, $\gamma_0$ the Lorentz factor of the injected plasma, $e$ and $m$ the positron charge and mass.
We fix $c=0.45$ cells/timestep, so that the temporal resolution of our simulations is $0.045\,\omega_{\rm{p}}^{-1}$ per timestep. At the final time $9000\;\omega_{\rm{p}}^{-1}$ of our simulation runs, the simulation domain is up to $\sim50000$ cells ($5000\,c/\omega_{\rm p}$) long along $x$. 

The incoming pair stream is injected along $-\mathbf{\hat{x}}$ with bulk Lorentz factor $\gamma_0=15$. The injected plasma is cold, with energy spectrum in the proper frame given by a three-dimensional Maxwellian  $f(\gamma)\propto\gamma\sqrt{\gamma^2-1}\exp(-\gamma/\Delta\gamma)$ with thermal spread $\Delta\gamma=10^{-4}$. Each computational cell is initialized with two electrons and two positrons. We also performed limited experiments with a larger number of particles per cell (up to 8 per species), obtaining essentially the same results. The magnetization of the upstream plasma is $\sigma=0.1$, where $\sigma\equiv B_{\rm u}^2/(4 \pi \gamma_0 mn_{\rm u} c^2)$ is the squared ratio of relativistic Larmor frequency $\Omega_{\msc{l}}\equiv eB_{\rm u}/(\gamma_0mc^2)$ to plasma frequency $\omega_{\rm p}$ for the injected flow ($B_{\rm u}$ is the upstream magnetic field measured in the wall frame). The upstream magnetization is chosen such that perpendicular shocks are not mediated by the relativistic \citet{weibel_59} instability.  As discussed by \citet{spitkovsky_05}, the growth of this instability is suppressed in perpendicular shocks if $\sigma\gtrsim10^{-3}$; as we show in \S\ref{sec:oblique}, for parallel shocks the Weibel instability may still be important for magnetizations as large as $\sigma=0.1$. The regime we explore may be relevant for PWNe and  internal shocks in GRBs and AGN jets, as we comment in \S \ref{sec:disc}. 

Keeping the magnetization fixed, we explore a range of magnetic inclinations by varying the angle $\theta$ between the shock direction of propagation $+\mathbf{\hat{x}}$ and the upstream magnetic field $\mathbf{B_{\rm u}}$, which is initialized to point towards the first quadrant of the $xz$ plane, as sketched in Fig.~\fig{simplane} (i.e., $B_{\rm {x,u}},B_{\rm {z,u}}\geq 0$ and $\theta\equiv \arctan[B_{\rm {z,u}}/B_{\rm {x,u}}]$). The magnetic obliquity angle $\theta$ is measured in the wall frame. We vary $\theta$ from $\theta=0^\circ$, which corresponds to a ``parallel'' shock, with magnetic field aligned with the shock normal, up to $\theta=90^\circ$, i.e., a ``perpendicular'' shock, with magnetic field along the shock front. For $\theta\neq 0^\circ$, in the upstream medium we also initialize a motional electric field $\mathbf{E_{\rm u}}=-\mbox{\boldmath{$\beta$}}_0\cross\mathbf{B_{\rm u}}$ along $-\mathbf{\hat{y}}$. Here, $\mbox{\boldmath{$\beta$}}_0=-\sqrt{1-1/\gamma_0^2}\;\bf{\hat{x}}$ is the three-velocity of the injected plasma. 

As a byproduct of the shock evolution, particles and electromagnetic waves may propagate upstream from the shock at the velocity of light. A fixed simulation box with a reflecting boundary on the left and an open boundary on the right would not be suitable to study the long-term evolution of the shock, since particles and fields would be removed from the simulation domain when they cross the open boundary, thus artificially suppressing any feedback they may have on the shock. Moreover, in the first stages of shock evolution,  physical quantities in a large portion of the simulation box have not been affected yet by the presence of the shock; following their time evolution would imply a waste of computational resources, in terms of both time and memory. Instead, we inject particles and fields into the simulation box from a ``moving injector'', which recedes from the wall (at $x_{\rm wall}=0$) along $+\mathbf{\hat{x}}$ at the speed of light, so that its position at time $t$ is roughly $x_{\rm inj}\sim c\,t$. This ensures that the simulation domain consists only of those regions which are in causal contact with the initial setup, thus saving on computational time. Also, our simulation box expands (via discrete jumps along $+\mathbf{\hat{x}}$) whenever the injector approaches its right boundary, thus allowing to save memory. In summary, a moving injector with an expanding domain permits to follow the shock evolution as far as the computational resources allow, preserving all the particles and waves generated by the shock. 

However, since the shock velocity is smaller than the light speed, the distance between the injector and the shock increases linearly with time. In all PIC codes, a numerical heating instability arises when cold relativistic plasma propagates for large distances over the numerical grid. The instability is a mixture of grid-Cerenkov instability -- due to the fact that electromagnetic waves on a discrete grid have subluminal phase velocity for large wavenumbers -- and spurious plasma oscillations (M.~Dieckmann, private communication). We reduce this numerical instability by using the fourth-order solver for Maxwell's equations \citep{greenwood_04}, which improves the numerical phase speed of the shortest wavemodes. To further get rid of the instability, we let our moving injector jump backwards (i.e., towards the wall) every 10000 time steps, resetting the fields to the right of the injector; in the following, we refer to this as a ``jumping injector''. The first jump happens after 40000 time steps and we tune the jump according to the shock speed so that the injected plasma never propagates farther than $1000\,c/\omega_{\rm{p}}$ from the injector before encountering the shock. Although this suppresses the growth of the numerical instability -- which arises if the injected plasma moves farther than $\sim1500\,c/\omega_{\rm p}$ from the injector -- it is not a completely satisfactory solution. In fact, each jump may remove from the computational domain some of the particles and waves propagating upstream, which would not happen if the injector were always receding at the speed of light. Also, electromagnetic waves moving away from the shock may be reflected back. We have tested the effects of a jumping injector by letting the first jump happen after 30000, 40000 and 50000 timesteps; in \S\ref{sec:time} we show that a jumping injector does not significantly affect our results.

\section{Shock structure}\label{sec:struct}
In this section we describe the structure and internal physics of relativistic magnetized pair shocks. We fix the upstream bulk Lorentz factor ($\gamma_0=15$) and magnetization ($\sigma=0.1$) and vary only the obliquity angle $\theta$, i.e., the angle between the upstream magnetic field and the shock normal, as measured in the wall frame. According to the criterion anticipated in \S\ref{sec:intro}, a shock is ``superluminal'' if particles cannot escape ahead of the shock by sliding along the magnetic field, and they are inevitably advected into the downstream medium. The critical boundary between subluminal and superluminal configurations should satisfy  in the \textit{upstream} fluid frame $\theta'_{\rm crit}=\arccos(\beta'_{\rm sh})$, where $\beta'_{\rm sh}$ is the upstream-frame shock speed, which itself depends on the magnetic obliquity. This corresponds to $\theta_{\rm crit}=\arctan[\gamma_0\tan(\theta'_{\rm crit})]$ in the \textit{wall} frame of our simulations. For $\gamma_0=15$ and $\sigma=0.1$, the shock velocity $\beta'_{\rm sh}$ computed from MHD equations yields $\theta_{\rm{crit}}\approx 34^\circ$ in the simulation frame, with weak dependence on both $\gamma_0$ and $\sigma$. Among the representative sample of magnetic obliquities that we present in this section, $\theta=0\deg$ (Figs.~\fig{fluid0} and \fig{subspace0}), $15\deg$ (Figs.~\fig{fluid15} and \fig{subspace15}) and $30\deg$ (Figs.~\fig{fluid30}  and \fig{subspace30}) are subluminal shocks, the last one being close to the critical obliquity angle $\theta_{\rm crit}\approx34\deg$; instead, $\theta=45^\circ$ (Figs.~\fig{fluid45} and \fig{subspace45}) is a representative superluminal shock. We anticipate (see \S\ref{sec:disc} for further discussion) that the main results presented below for $\gamma_0=15$ and $\sigma=0.1$ hold across a larger range of upstream Lorentz factors (we tried $\gamma_0=5$ and $\gamma_0=50$, for fixed $\sigma=0.1$) and magnetizations (we tested $\sigma=0.03$ and $\sigma=0.3$, for fixed $\gamma_0=15$).

Figs.~\fig{fluid0}, \fig{fluid15}, \fig{fluid30} and \fig{fluid45} present the shock internal structure as a function of the longitudinal coordinate $x$  at time $\ompt=2250$. We include both two-dimensional plots in the $xy$ simulation plane and transversely-averaged (along $y$) plots for the following quantities: particle number density (panel \textit{a} for the two-dimensional plot and panel \tit{b} for the corresponding $y$-average), magnetic energy fraction $\epsilon_{\msc{b}}$ (defined as the ratio of magnetic energy density to kinetic energy density of the incoming flow, panels \textit{c} and \tit{d}), transverse magnetic field $B_{\rm z}$ (panels \textit{e} and \tit{f}), transverse electric field $E_{\rm y}$ (panels \textit{g} and \tit{h}), longitudinal electric field $E_{\rm x}$ (panels \textit{i} and \tit{j}). We also plot the $y$-averaged electric pseudo-potential $-\int \langle E_{\rm x}\rangle\, \ud x$ in panel \textit{k}, normalized to the kinetic energy of injected particles. The positron phase space density projected onto the $x-\gb{x}$, $x-\gb{y}$ and $x-\gb{z}$ planes is shown in panels \textit{l,\,m,\,n} respectively. Figs.~\fig{subspace0},\,\fig{subspace15},\,\fig{subspace30},\,\fig{subspace45} show the positron $\gb{x}-\gb{z}$ momentum space  and electron and positron energy spectra at three different locations through the flow, as marked by arrows at the bottom of Figs.~\fig{fluid0},\,\fig{fluid15},\,\fig{fluid30},\,\fig{fluid45} respectively.

In all the simulation runs, a collisionless shock is formed which propagates away from the wall. The jump in density and electromagnetic fields at the shock, as well as the shock velocity, are in agreement with MHD calculations, as we show in \S\ref{sec:oblique}. For oblique configurations ($0\deg<\theta<90\deg$, i.e., with the exception of parallel and perpendicular shocks), the magnetic field refracts at the shock. In the frame of the simulation, the longitudinal speed of the incoming flow drops to zero after the shock; however, for oblique configurations, the downstream fluid preserves a small transverse velocity $\beta_{\rm z,d}<0$, in agreement with MHD predictions. In this respect, the wall frame of our simulations does not strictly coincide with the downstream fluid frame, but for most purposes $\beta_{\rm z,d}$ is small enough that the two frames are almost undistinguishable. In the shock transition layer, the incoming cold beam isotropizes and thermalizes to a downstream temperature that relates to the upstream kinetic energy, reduced by the fraction of upstream particle energy converted to downstream magnetic energy. For low obliquities ($\theta\lesssim45\deg$), the downstream plasma is fully isotropic in three dimensions. For larger obliquities, the plane perpendicular to the downstream magnetic field, where particle orbits would lie in the absence of significant motion along the field, is almost degenerate with the direction of propagation of the incoming plasma. In this case, particle motions will mostly be confined to the plane orthogonal to the downstream field, which prevents efficient isotropization along the direction of the field. In other words, for low obliquities the effective adiabatic index of the downstream plasma is appropriate for a three-dimensional relativistic gas ($\Gamma=4/3$), whereas for $\theta\gtrsim45\deg$ the adiabatic index tends to that of a two-dimensional relativistic fluid ($\Gamma=3/2$). 

This section is structured as follows. In \S\ref{sec:oblique} we present a general description of the  structure of oblique shocks. We discuss that Weibel-like filamentation instabilities are likely to mediate the shock formation for low magnetic obliquities, whereas magnetic reflection from the shock-compressed magnetic field triggers high-obliquity shocks. For $0\deg<\theta<90\deg$, the longitudinal profile of particle density, magnetic energy and transverse magnetic field shows a fluid structure with two shocks, a strong ``fast'' shock (which we identify with the main shock) and a weak ``slow'' shock. The velocity and jump conditions of the fast shock are in agreement with MHD calculations. 

In \S \ref{sec:sub} and \S\ref{sec:super} we describe in detail some representative examples of subluminal ($0\leq\theta\lesssim\theta_{\rm crit}$) and superluminal ($\theta_{\rm crit}\lesssim\theta\leq90\deg$) shocks. In subluminal configurations, some shock-accelerated particles escape ahead of the shock, and they populate a diffuse beam of high-energy  particles propagating upstream. These ``returning'' particles trigger the generation of upstream waves, which propagate towards the shock with a wavevector oblique to the background magnetic field. In superluminal shocks, no returning high-energy particles or upstream waves are observed. The importance of these oblique modes for particle acceleration will be discussed in \S\ref{sec:mechanism}, where we present a detailed investigation of the acceleration mechanisms for subluminal shocks.

\subsection{Oblique shocks: general overview}\label{sec:oblique}
\begin{figure*}[tbp]
\unitlength = 0.0011\textwidth
\begin{center}
\begin{picture}(550,700)(0,0)
\includegraphics[width=.6\textwidth]{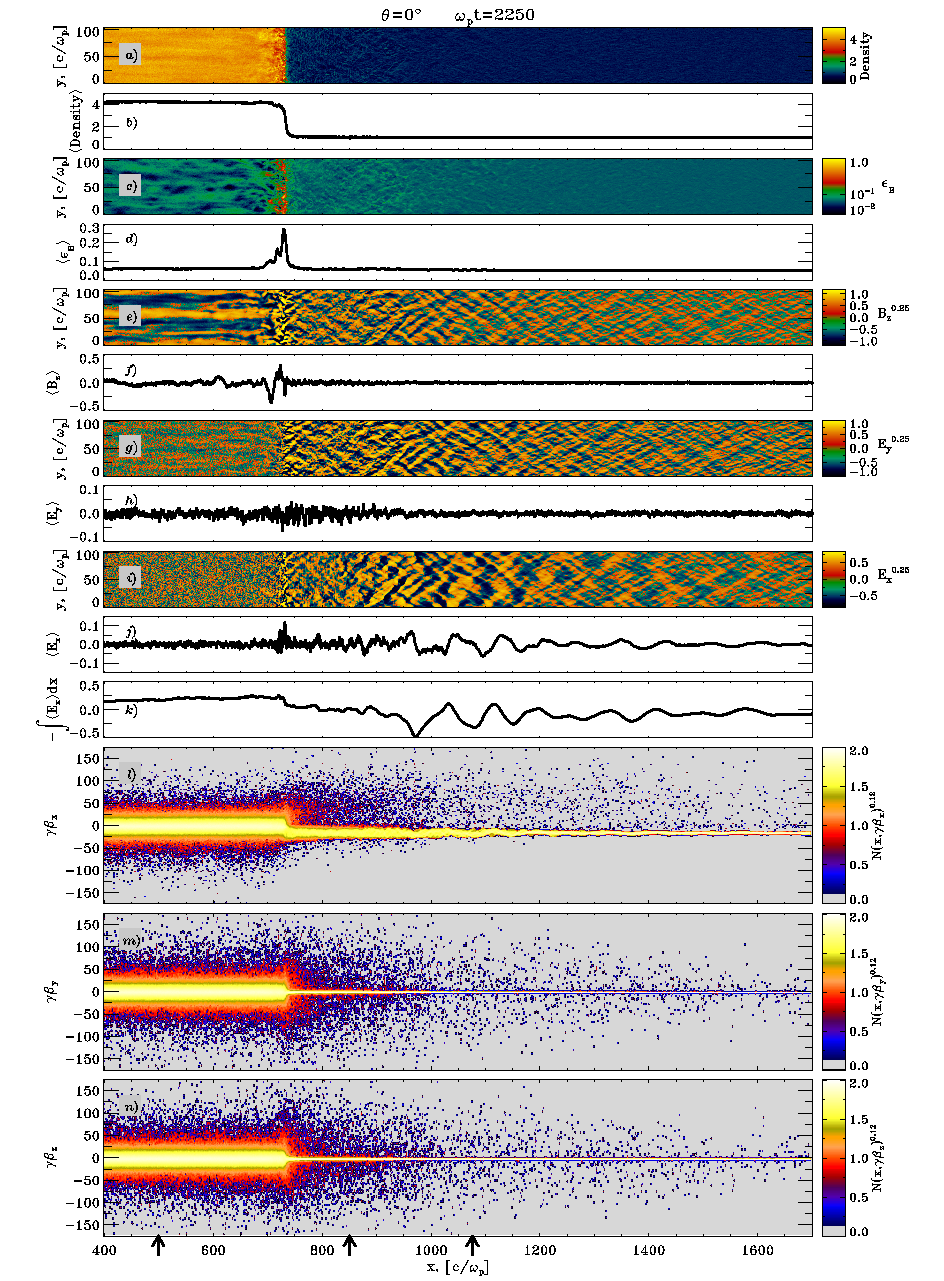}
\end{picture}
\caption{Shock structure and positron phase space at time $\omega_{\rm{p}}t=2250$ for $\theta=0^\circ$: \tit{a}) Particle number density in the simulation plane, normalized to the upstream value; \tit{b}) Transversely-averaged number density; \tit{c}) Magnetic energy fraction $\epsilon_{\msc{b}}$, i.e., the ratio of magnetic energy density to the kinetic energy density of the incoming flow; \tit{d}) Transversely-averaged $\langle \epsilon_{\msc{b}}\rangle$; \tit{e})-\tit{f}) $B_{\rm z}$ and transversely-averaged $\langle B_{\rm z}\rangle$, in units of the upstream background magnetic field; \tit{g})-\tit{h}) $E_{\rm y}$ and transversely-averaged $\langle E_{\rm y}\rangle$, in units of the upstream background magnetic field; \tit{i})-\tit{j}) $E_{\rm x}$ and transversely-averaged $\langle E_{\rm x}\rangle$, in units of the upstream background magnetic field; \tit{k}) Electric pseudo-potential $-\int \langle E_{\rm x}\rangle\, \ud x$, normalized to the kinetic energy of incoming particles; \tit{l})-\tit{n}) Longitudinal phase space plots $x-\gamma\beta_{i}$ ($i=x,\,y,\,z$) of positrons, with phase space density plotted as a two-dimensional histogram. For the two-dimensional plots of $B_{\rm z}$, $E_{\rm y}$ and $E_{\rm x}$, $q^\alpha$ stands for $(q/|q|)|q|^\alpha$. }
\label{fig:fluid0}
\end{center}
\end{figure*}

\begin{figure*}
\unitlength = 0.0011\textwidth
\begin{center}
\begin{picture}(550,281)(0,0)
\includegraphics[width=.6\textwidth]{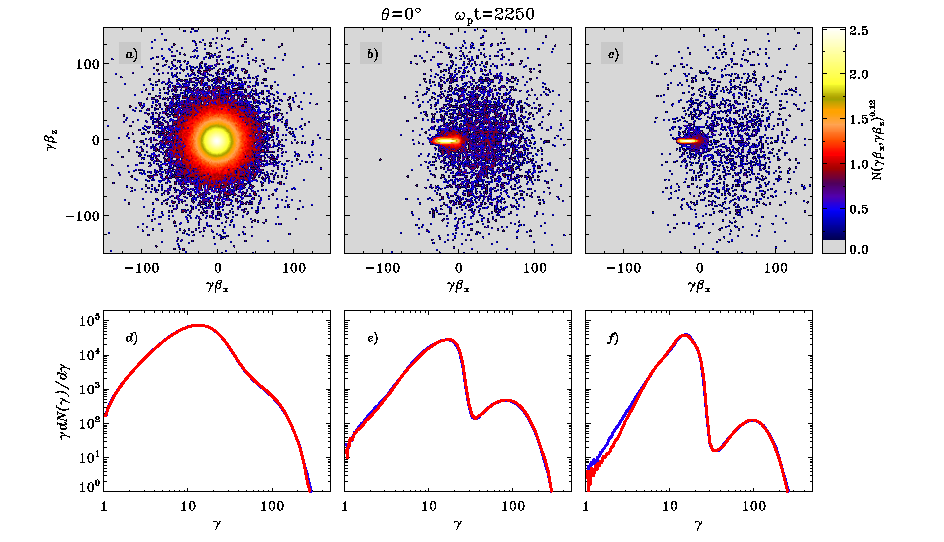}
\end{picture}
\caption{Positron momentum space and electron (blue) and positron (red) spectra at time $\omega_{\rm{p}}t=2250$ for $\theta=0^\circ$, at three slices through the flow as marked by arrows at the bottom of Fig.~\fig{fluid0}.  \tit{a})-\tit{c}) Momentum space $\gamma\beta_{\rm x}-\gamma\beta_{\rm z}$ in the three slabs, with momentum space density plotted as a two-dimensional histogram; \tit{d})-\tit{f}) Electron (blue) and positron (red) spectra $\gamma\,\ud  N/\ud\gamma$ in the three slices, normalized to the total number of particles within the leftmost slab (apart from a constant multiplication factor).}
\label{fig:subspace0}
\end{center}
\end{figure*}

Unmagnetized relativistic shocks are mediated by Weibel-like filamentation instabilities \citep{weibel_59, medvedev_loeb_99, gruzinov_waxman_99}. Incoming particles are scattered by the Weibel-generated fields and the incoming flow is isotropized, decelerated and compressed \citep[e.g.,][]{spitkovsky_05}. The same mechanism is likely to mediate moderately magnetized ($\sigma\lesssim1$) parallel shocks. In fact, we find that the width of the transition layer in $\sigma=0.1$ parallel shocks, as evaluated from the density profile (Fig.~\fig{fluid0}\textit{\,b} at $x_{\rm sh}\approx730\comp$), is of the order of a few tens of skin depths, as observed in unmagnetized shocks.\footnote{We point out that the shock transition region in strongly magnetized ($\sigma\gtrsim1$) plasmas is much wider, since cross-field motion is suppressed and the Weibel instability is quenched.}
Also, the magnetic energy density peak at the shock (reaching $\sim30\%$ of the upstream kinetic energy density, see Fig.~\fig{fluid0}\textit{\,d}) is a characteristic signature of magnetic field amplification by a Weibel-like process. The field then decays to the plateau predicted by MHD calculations on a length scale $\sim50\comp$.

Oblique shocks ($\theta\neq0\deg$) may be triggered by another process, namely magnetic reflection of the incoming flow from the shock-compressed magnetic field \citep{alsop_arons_88}, as reported for perpendicular shocks by previous one-dimensional \citep[e.g.,][]{langdon_88,gallant_92} and three-dimensional \citep{spitkovsky_05} PIC simulations. The synchrotron maser instability, excited by coherent gyration of the incoming particles in the shock magnetic field, mediates the thermalization of the upstream plasma \citep{hoshino_91}. Indeed, the peak in the transversely-averaged density profile observed for  $\theta=45\deg$ at the shock location (see Fig.~\fig{fluid45}\textit{\,b} at $x_{\rm sh}\approx920\comp$)  is due to  Larmor gyration of the incoming particles in the compressed shock fields, which decelerates the upstream flow and therefore increases its density. The density rise is as sharp as a few Larmor radii in the compressed field (i.e., a few plasma skin depths), much thinner than for parallel shocks (compare Fig.~\fig{fluid45}\textit{\,b} with Fig.~\fig{fluid0}\textit{\,b}). In just a few Larmor scales downstream from the shock, the density profile flattens to a plateau in reasonable agreement with MHD jump conditions. Shocks with larger obliquities ($\theta\gtrsim45\deg$) are also triggered by magnetic reflection.

\begin{figure*}[tbp]
\begin{center}
\includegraphics[width=0.6\textwidth]{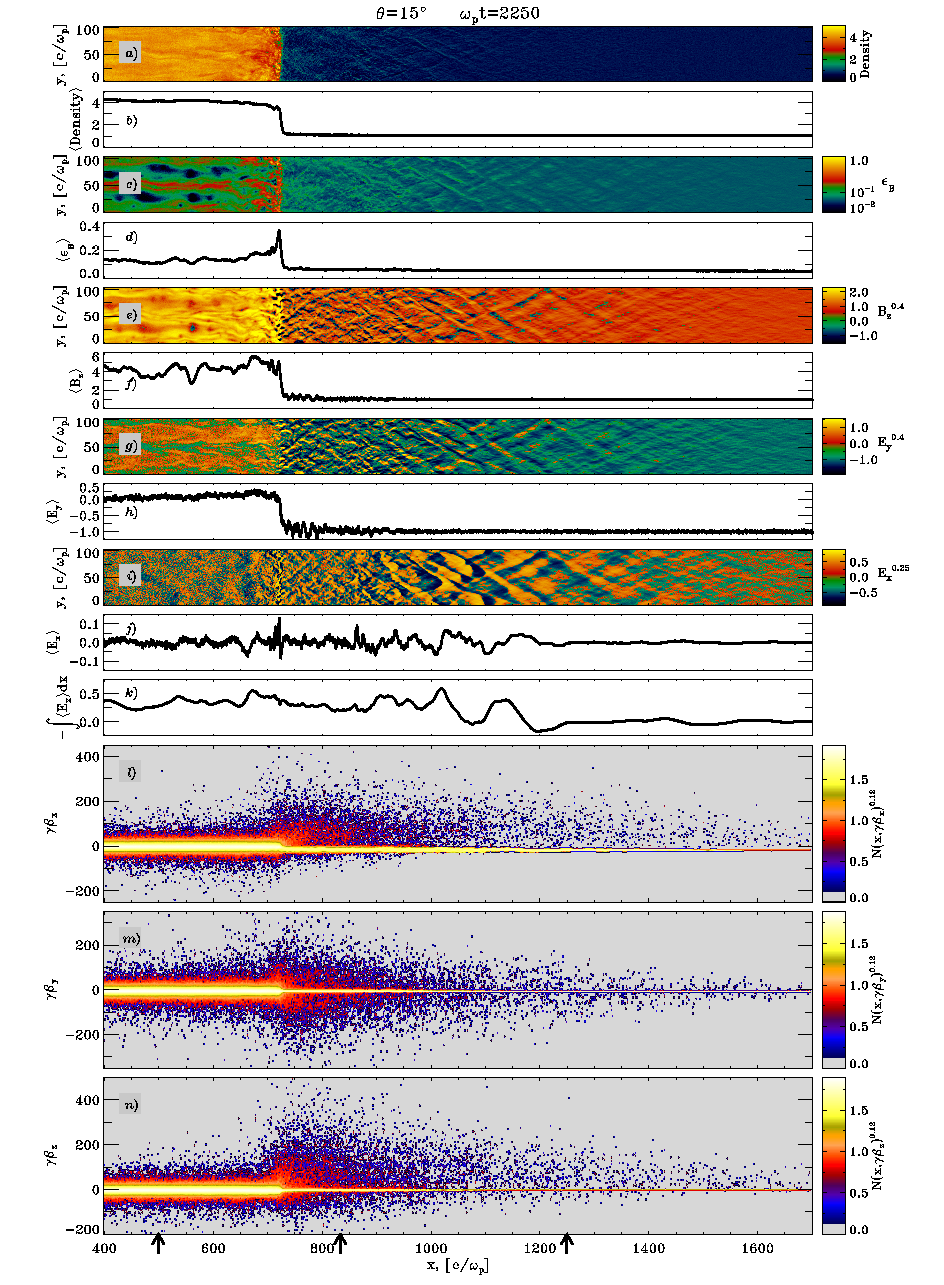}
\caption{Shock structure and positron phase space at time $\omega_{\rm{p}}t=2250$ for $\theta=15^\circ$. The fluid quantities are normalized as in Fig.~\fig{fluid0}, but here $B_{\rm z}$ and $E_{\rm y}$ are in units of the transverse component $B_{\rm{z,u}}$ of the upstream background magnetic field. The excess of magnetic energy at the shock (panel \tit{d} at $x_{\rm sh}\approx720\comp$) does not show up in the transversely-averaged $\langle B_{\rm z}\rangle$ (panel \tit{f}) due to cancellation between magnetic fluctuations of opposite sign.}
\label{fig:fluid15}
\end{center}
\end{figure*}

\begin{figure*}[tbp]
\begin{center}
\includegraphics[width=0.6\textwidth]{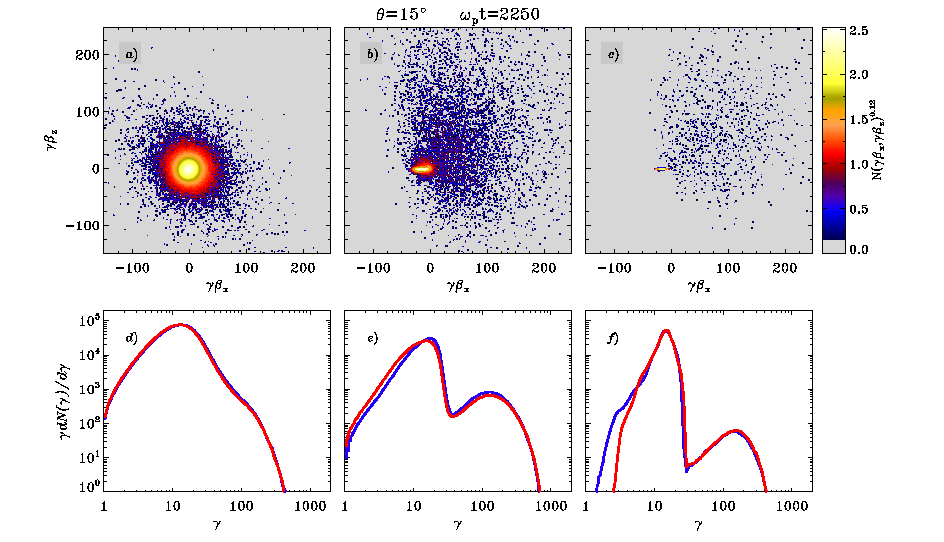}
\caption{Positron momentum space and electron (blue) and positron (red) spectra at time $\omega_{\rm{p}}t=2250$ for $\theta=15^\circ$, at three slices through the flow as marked by arrows at the bottom of Fig.~\fig{fluid15}. See the caption of Fig.~\fig{subspace0} for details.}
\label{fig:subspace15}
\end{center}
\end{figure*}

\begin{figure*}[tbp]
\begin{center}
\includegraphics[width=0.6\textwidth]{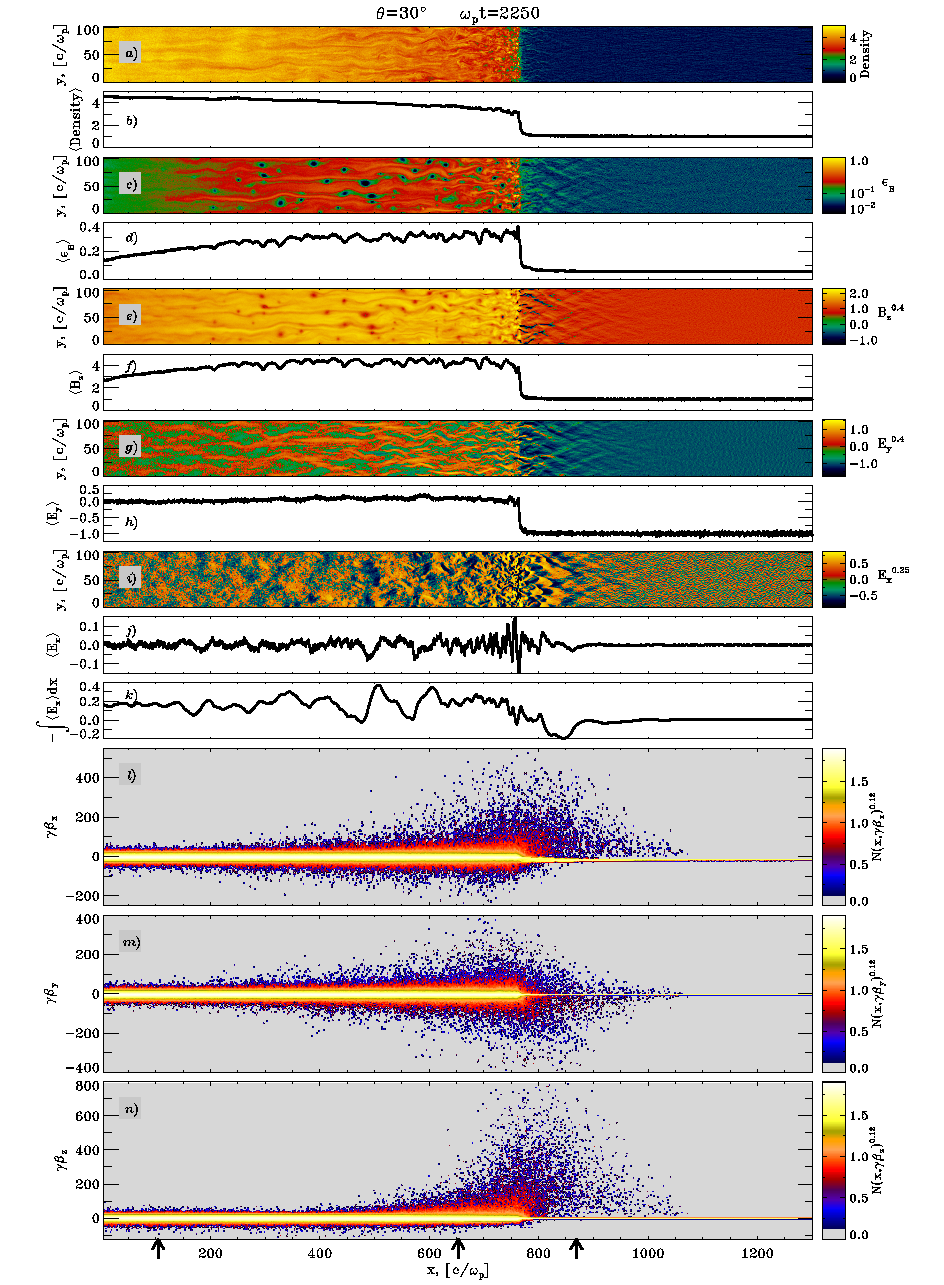}
\caption{Shock structure and positron phase space at time $\omega_{\rm{p}}t=2250$ for $\theta=30^\circ$. The fluid quantities are normalized as in Fig.~\fig{fluid15}. }
\label{fig:fluid30}
\end{center}
\end{figure*}

\begin{figure*}[tbp]
\begin{center}
\includegraphics[width=0.6\textwidth]{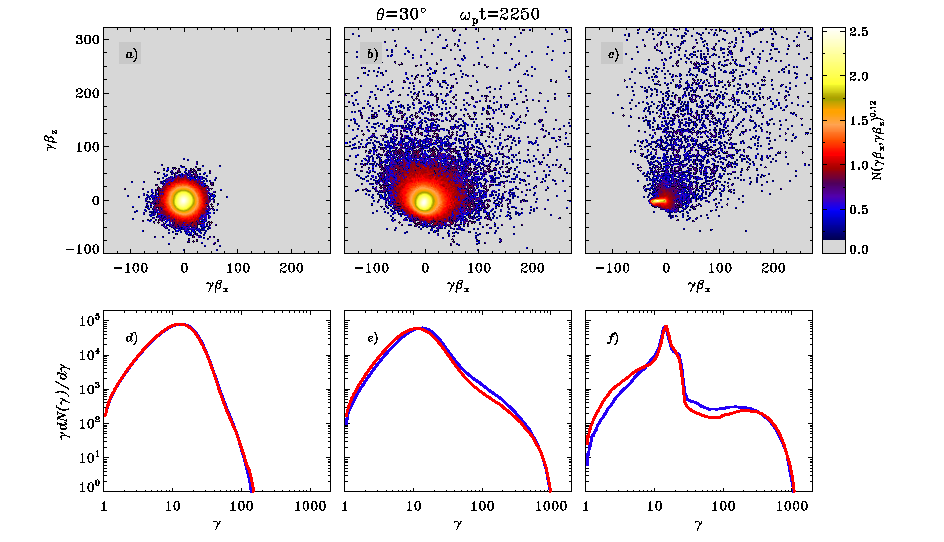}
\caption{Positron momentum space and electron (blue) and positron (red) spectra at time $\omega_{\rm{p}}t=2250$ for $\theta=30^\circ$, at three slices through the flow as marked by arrows at the bottom of Fig.~\fig{fluid30}.  See the caption of Fig.~\fig{subspace0} for details.}
\label{fig:subspace30}
\end{center}
\end{figure*}

\begin{figure*}[!htbp]
\begin{center}
\includegraphics[width=0.6\textwidth]{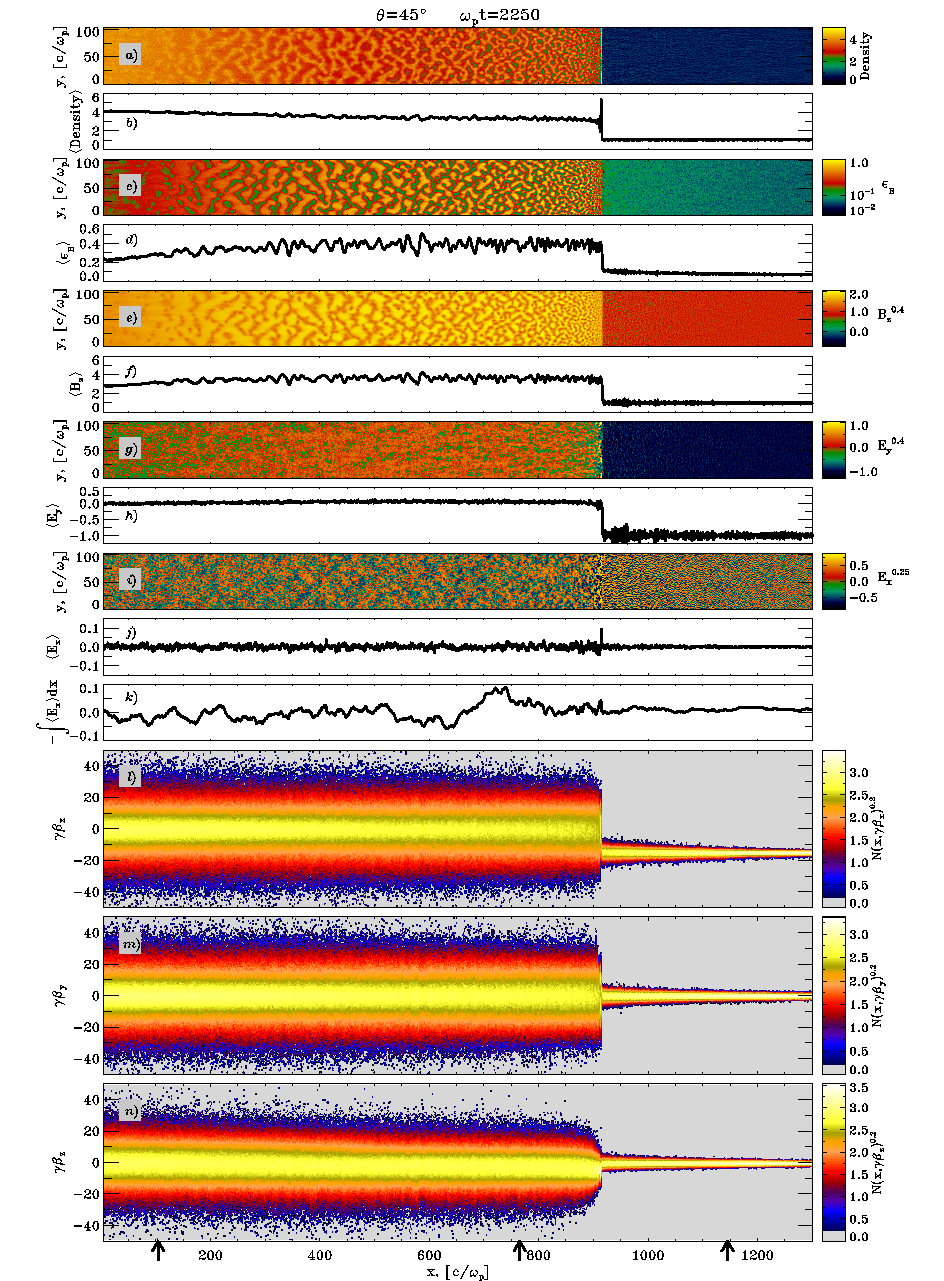}
\caption{Shock structure and positron phase space at time $\omega_{\rm{p}}t=2250$ for $\theta=45^\circ$. The fluid quantities are normalized as in Fig.~\fig{fluid15}. }
\label{fig:fluid45}
\end{center}
\end{figure*}

Both Weibel instabilities and magnetic reflection play a role in mediating shocks with intermediate obliquities ($0\deg<\theta\lesssim45\deg$). Filamentation instabilities dominate for lower obliquities, such as $\theta=15\deg$, whose fluid structure resembles $\theta=0\deg$ in the magnetic overshoot at the shock (Fig.~\fig{fluid15}\pan{d} at $x_{\rm sh}\approx720\comp$). Instead, such Weibel-induced magnetic energy excess is not seen for $\theta=30\deg$ (Fig.~\fig{fluid30}\pan{d} at $x_{\rm sh}\approx770\comp$) or larger obliquities; rather, investigation of early times ($\omega_{\rm{p}}t=225$) for $\theta=30\deg$ shows that this shock started out  with the prominent density peak characteristic of highly-oblique shocks, which suggests that this shock is mainly triggered by magnetic reflection. For $\theta=30\deg$, the sharp density peak observed at early times was later smoothed by the increased pressure in shock-accelerated particles.

For all oblique magnetic configurations ($0^\circ<\theta<90^\circ$), a fluid structure with two shocks (a strong ``fast'' shock and a weak ``slow'' shock) is seen in our PIC simulations. The presence of two shocks is predicted for the parameters ($\gamma_0=15$ and $\sigma=0.1$) that we adopt \citep{landau_60} and has been confirmed by MHD simulations of the collision of a magnetized shell against a wall (S.~Komissarov, private communication). The particle number density rises across both shocks; however, while the fast shock involves an amplification of the magnetic field, across the slow shock the magnetic energy decreases. The slow shock propagates into the downstream medium of the fast shock. Its transition layer in our PIC simulations is not as narrow as MHD would predict, due to insufficient dissipation along the magnetic field. Rather, the slow-shock transition region stretches self-similarly with time; the jump in fluid quantities between the wall and the fast-shock downstream stays roughly constant in time. We observe that the slow shock is slower and weaker as $\theta$ approaches either $0^\circ$ or $90^\circ$, and relatively stronger and faster for intermediate obliquities ($\theta\approx30\deg$). For oblique magnetic configurations, we shall refer to the fast shock simply as ``the shock'' and to its downstream region as ``the downstream medium''.

For the main (i.e., fast) shock, the velocity and jump conditions extracted from our PIC simulations agree within a few percent with the solution of relativistic MHD equations \citep{appl_camenzind_88} supplemented by the \textit{three-dimensional} \citet{synge_57} equation of state, provided that MHD results are Lorentz-transformed from the standard shock frame into the simulation frame.\footnote{Shock speed and jump conditions from our PIC simulations are computed at relatively early times ($\omega_{\rm{p}}t=2250$), since for later stages of the shock evolution the pressure by accelerated particles may significantly affect the shock structure.} In Table \ref{tab1} we compare the results from PIC simulations and MHD calculations for the shock velocity $\beta_{\rm sh}$ and the jump across the shock in particle number density ($n_{\rm d}/n_{\rm u}$), magnetic energy density ($B^2_{\rm d}/B^2_{\rm u}$), transverse magnetic field ($B_{\rm z,d}/B_{\rm z,u}$) and transverse electric field ($E_{\rm y,d}/E_{\rm y,u}$). We remark that, for oblique shocks ($0^\circ<\theta<90^\circ$), the background motional electric field (last column in Table \ref{tab1}) does not drop to zero after the shock since, as discussed above, the downstream flow preserves a small transverse velocity $\beta_{\rm z,d}<0$ which gives a motional field $E_{\rm y,d}=-\beta_{\rm z,d}B_{\rm x,d}>0$. 

As shown in Table \ref{tab1}, for $\theta\lesssim30^\circ$ our kinetic calculations are in remarkable agreement with \tit{three-dimensional} MHD results, suggesting that for low magnetic obliquities the downstream plasma in our \tit{two-dimensional} simulation box has a \tit{three-dimensional} effective adiabatic index, as anticipated above. For $\theta\gtrsim45\deg$, the adiabatic index of the downstream fluid in our simulations tends to that of a two-dimensional gas, which explains the discrepancy between PIC simulations and three-dimensional MHD calculations for $\theta=45\deg$ in Table \ref{tab1}.\footnote{We point out that the downstream adiabatic index in high-obliquity shocks tends to that of a two-dimensional fluid also in \tit{three-dimensional} PIC simulations, since particle isotropization along the field is still harder than orthogonal to the field. However, in our two-dimensional runs,  the simulation plane for high obliquities is almost degenerate with the plane orthogonal to the downstream field, where particle gyro-orbits lie, and this makes such isotropization even harder.} 

\begin{table}[tbhp]
\caption{Comparison between PIC simulations (upper row) and MHD calculations (lower row)}
\centering
\begin{tabular}{|c|c|c|c|c|c|}
\hline
$\theta$ & $\beta_{\rm{sh}}$ & $n_{\rm{d}}/n_{\rm{u}}$ & $B^{2^{\phantom{1}}}_{\rm{d}}/B^{2^{\phantom{1}}}_{\rm{u}}$ & $B_{\rm{z,d}}/B_{\rm{z,u}}$ & $E_{\rm{y,d}}/E_{\rm{y,u}}$\\[0.8ex] 
\hline
\hline 
$0^\circ$  & 0.32 & 4.1 & 1.1 & &\\& 0.31 & 4.2 & 1.0 & &\\
\hline 
$15^\circ$ & 0.32 & 4.0 & 2.4 & 4.3 & -0.08\\& 0.32 & 4.1 & 2.3 & 4.5 & -0.16\\
\hline 
$30^\circ$  & 0.34 & 3.8 & 5.2 & 4.3 & -0.10\\& 0.35 & 3.9 & 5.1 & 4.2 & -0.11\\
\hline 
$45^\circ$ & 0.41 & 3.2 & 6.5 & 3.6 & -0.06\\& 0.37 & 3.7 & 7.8 & 3.8 & -0.07\\
\hline
\end{tabular}
\label{tab1}
\vspace{2mm}
\end{table}

Another common feature of all oblique and perpendicular shocks ($0^\circ<\theta\leq90^\circ$) is the presence of a transient electromagnetic precursor (seen in $B_{\rm z}$ and $E_{\rm y}$ for the magnetic configuration shown in Fig.~\fig{simplane}) propagating at the speed of light ahead of the shock, as reported in previous one-dimensional \citep[e.g.,][]{gallant_92, hoshino_92, hoshino_08} and short three-dimensional \citep{spitkovsky_05} PIC simulations of perpendicular shocks. This is a transverse electromagnetic wave which is generated in the initial stages of shock evolution by bunching of coherent Larmor orbits near the shock front. In multi-dimensional simulations, we observe that it is no longer produced as soon as the coherence between different locations along the shock surface is lost. The transient precursor generated at early times survives as  
a dispersive wave packet propagating upstream, farther from the shock than Figs.~\fig{fluid15},\,\fig{fluid30},\,\fig{fluid45} show. 
We find that its power increases with  magnetic obliquity and its Poynting flux may be comparable or in excess of $c\,B_{\rm z,u}^2/4 \pi$. 

\subsection{Subluminal shocks: $0\leq\theta\lesssim\theta_{\rm crit}$}\label{sec:sub}
The internal structure at time $\ompt=2250$ of shocks with magnetic obliquities $\theta=0\deg$ (Figs.~\fig{fluid0},\,\fig{subspace0}), $15\deg$ (Figs.~\fig{fluid15},\,\fig{subspace15}) and $30\deg$ (Figs.~\fig{fluid30},\,\fig{subspace30}) shows some features common to all subluminal magnetic configurations. Subluminal shocks are characterized by the presence of a diffuse stream of high-energy shock-accelerated particles that propagate ahead of the shock and trigger the growth of oblique waves in the upstream medium. In turn, as we explain in \S\ref{sec:mechanism}, such waves are of fundamental importance for particle acceleration.

The longitudinal phase space plots of positrons in panels \textit{l-n} of Figs.~\fig{fluid0},\,\fig{fluid15},\,\fig{fluid30} show the injected plasma as a cold dense beam propagating with $\gamma\beta_{\rm x}\approx-15$ and broadening while it approaches the shock, which is located at $x_{\rm sh}\approx730\comp$ for $\theta=0\deg$, at $x_{\rm sh}\approx720\comp$ for $\theta=15\deg$ and at $x_{\rm sh}\approx770\comp$ for $\theta=30\deg$. In the shock transition layer, the incoming plasma is isotropized and thermalized. A population of  high-energy particles moving ahead of the shock appears in the upstream region as a hot diffuse beam with $\gamma\beta_{\rm x}>0$ (panel \tit{l} in Figs.~\fig{fluid0},\,\fig{fluid15},\,\fig{fluid30}); as we show in \S\ref{sec:mechanism}, these particles have been accelerated at the shock. Additional insight is provided by the positron $\gb{x}-\gb{z}$  momentum space and electron (blue) and positron (red) spectra in Figs.~\fig{subspace0},\,\fig{subspace15},\,\fig{subspace30}, extracted from three different locations across the flow as  marked by arrows at the bottom of Figs.~\fig{fluid0},\,\fig{fluid15},\,\fig{fluid30} respectively. Upstream slabs (panels \tit{c} and \tit{f} in Figs.~\fig{subspace0},\,\fig{subspace15},\,\fig{subspace30}) show two components: the cold incoming plasma appears in particle spectra (panel \tit{f} in Figs.~\fig{subspace0},\,\fig{subspace15},\,\fig{subspace30}) as a relatively narrow bump peaking at low energies, around $\gamma\approx15$, which broadens as the flow approaches the shock; the returning diffuse beam of shock-accelerated particles  populates the high-energy bump, which is more and more depleted (especially at intermediate energies) as we move farther upstream. Far downstream slabs (panels \tit{a} and \tit{d} in Figs.~\fig{subspace0},\,\fig{subspace15},\,\fig{subspace30}) show roughly isotropic particle distributions with  thermal spectra supplemented by high-energy tails. For $\theta=30\deg$, the downstream population of energetic particles significantly varies along $x$, with regions closer to the shock showing more particles of higher energies than slices farther downstream (Fig.~\fig{fluid30}\pan{l-n}, and compare panels \tit{a},\pan{d} with panels \tit{b},\pan{e} in Fig.~\fig{subspace30}). Since fluid elements which are farther downstream passed through the shock at earlier times, this suggests that the shock acceleration efficiency increased over time, as we discuss in \S\ref{sec:time}.

For oblique magnetic configurations (Fig.~\fig{fluid15} for $\theta=15\deg$ and Fig.~\fig{fluid30} for $\theta=30\deg$), the phase space plots $x-\gb{y}$ (panel \tit{m}) and $x-\gb{z}$ (panel \tit{n}) are not symmetric around $\gamma\beta_{i}=0$ ($i=y,\,z$), when focusing on the high-energy particles located in the vicinity of the shock (see also panels \tit{b},\,\tit{c} in Figs.~\fig{subspace15},\,\fig{subspace30}). The high-energy positrons with $\gamma\beta_{\rm z}>0$ in the shock and upstream regions are shock-accelerated particles whose gyro-center is heading upstream along the oblique magnetic field (we remind that $B_{\rm x}, B_{\rm z}>0$).\footnote{For $\theta=30\deg$, the $x$-velocity of a particle sliding at the speed of light along the upstream magnetic field would be $\approx0.5\,c$ in the simulation frame. This is compatible with the longitudinal location $x\approx1100\comp$ at time $\ompt=2250$ for the returning particles which are farthest upstream (Fig.~\fig{fluid30}\pan{l-n}). The same is true for smaller obliquities, but the leading edge of the population of returning particles is outside the $x$-range plotted in Figs.~\fig{fluid0} and \fig{fluid15}.} 
With increasing magnetic obliquity from $\theta=15\deg$ to $30\deg$, the asymmetry around $\gb{z}=0$ becomes more evident and the returning high-energy positrons show increasingly larger $\gb{z}$ than $\gb{x}$ (compare panels \tit{l} and \tit{n} between Fig.~\fig{fluid15} and Fig.~\fig{fluid30}). The electron phase space $x-\gb{z}$ (not plotted) shows the same asymmetry. 

The excess of high-energy positrons with $\gb{y}<0$ observed in the vicinity of the shock (panel \tit{m} of Figs.~\fig{fluid15} and \fig{fluid30}) is a natural consequence of the $\grad B$ drift experienced by particles accelerated at the shock. For positively-charged particles, the drift is directed along $-\hat{\mathbf{y}}$, which explains the sign of the asymmetry in Figs.~\fig{fluid15}\textit{\,m},\,\fig{fluid30}\textit{\,m}.  Since the magnetic field jump at the shock is larger for increasing obliquities (Table \ref{tab1}), and the drift velocity is then higher, we expect the asymmetry to be more significant for $\theta=30\deg$ than for $15\deg$, as observed. Moreover, since electrons and positrons move in opposite directions under $\grad B$ drift, the electron $x-\gb{y}$ phase space (not shown) exhibits a reversed asymmetry, with an excess of particles having $\gamma\beta_{\rm y}>0$.

The returning beam of high-energy particles triggers in the upstream medium the generation of waves whose wavevector is oblique to the background magnetic field. Such waves have an electromagnetic component ($B_{\rm z}$ in panel \tit{e}  and $E_{\rm y}$ in panel \tit{g} of Figs.~\fig{fluid0},\,\fig{fluid15},\,\fig{fluid30}) and a smaller electrostatic component ($E_{\rm x}$ in panel \tit{i} of Figs.~\fig{fluid0},\,\fig{fluid15},\,\fig{fluid30}). The electrostatic component also appears as upstream oscillations in the transversely-averaged $\langle E_{\rm x}\rangle$ (panel \tit{j} of Figs.~\fig{fluid0},\,\fig{fluid15},\,\fig{fluid30}) and in the electric pseudo-potential $-\int \langle E_{\rm x}\rangle\, \ud x$ (panel \tit{k} of Figs.~\fig{fluid0},\,\fig{fluid15},\,\fig{fluid30}). The upstream waves grow in amplitude while propagating towards the shock, where the regular oblique pattern observed far upstream (with characteristic wavelength $\sim20\,c/\omega_{\rm p}$ at time $\omega_{\rm{p}}t=2250$) changes to a more complicated structure (e.g., compare $x\gtrsim1000\comp$ with $750\comp\lesssim x\lesssim900\comp$ for $\theta=15\deg$ in Fig.~\fig{fluid15}\pan{e},\pan{g},\pan{i}). The ratio between the magnetic energy of the oblique waves (mostly contributed by their $B_{\rm z}$) and of the background upstream field is of the order of a few tens of percent, roughly constant with magnetic obliquity. 

By comparing panels \tit{e},\pan{g},\pan{i} with panels \tit{l-n} of Fig.~\fig{fluid30} for $\theta=30\deg$, the growth of the upstream oblique modes appears to be spatially coincident (at $x\approx1000\comp$) with the leading edge of the population of returning particles. The same holds true for smaller obliquities, but for $\theta=0\deg$ and $15\deg$ the returning particles that are farthest upstream are outside the $x$-range plotted in Figs.~\fig{fluid0} and \fig{fluid15}. This suggests that interplay between the unperturbed incoming plasma and the returning high-energy beam may be responsible for triggering the oblique waves.   

The growth of the upstream waves is neither an unphysical consequence of our jumping injector nor an effect of the numerical heating associated with the propagation of a cold beam through the grid. In fact, if we replace the reflecting wall at $x_{\rm wall}=0$ with an open boundary, everything else being equal, we observe that  no upstream waves are produced, nor is there a shock. In order to test the role played by the returning high-energy particles in generating the upstream waves, we have performed a suite of simulations in which we artificially suppress particle acceleration. In these runs, we introduced cooling, so that particles with $\gamma>\gamma_{\rm cool}$ lose a random fraction of their nonthermal energy \citep[a similar technique was used in][]{keshet_08}. By varying $\gamma_{\rm cool}$ (we tried $\gamma_{\rm cool}=50,\,80,\,140$ for $\theta=0\deg$), we conclude that particle cooling suppresses the growth of the upstream waves, and by decreasing $\gamma_{\rm cool}$ (i.e., by forcing more particles to cool) both the wave amplitude and the distance from the shock where they are triggered decrease. 

In summary, the oblique waves seen in the upstream medium are generated by the returning beam of high-energy particles; they may result from  a mixed-mode streaming instability which couples the electrostatic two-stream instability with the electromagnetic filamentation (Weibel) instability, as envisioned by \citet{bret_08}. A complete analysis of the wave dispersion relation and generation mechanism is deferred to a future paper.

Finally, we point out that the electrostatic component of the upstream waves induces oscillations in the \mbox{$\gb{x}$-velocity} of the incoming plasma, as seen in the positron phase space of Fig.~\fig{fluid0}\pan{l}. The electron $x-\gb{x}$ phase space (not shown) presents oscillations of opposite sign. The difference at low energies between electron and positron spectra in the upstream medium (panel \tit{f} in Figs.~\fig{subspace0},\,\fig{subspace15},\,\fig{subspace30}) can be correlated with the sign of the electric pseudo-potential at the same longitudinal location (panel \tit{k} in Figs.~\fig{fluid0},\,\fig{fluid15},\,\fig{fluid30} respectively), since a net electric potential accelerates incoming charges of one sign and decelerate charges of the opposite sign. We further comment on the effects on particle spectra of the electrostatic component of upstream oblique waves  in \S \ref{sec:time}.

\begin{figure*}[!hbtp]
\begin{center}
\includegraphics[width=0.6\textwidth]{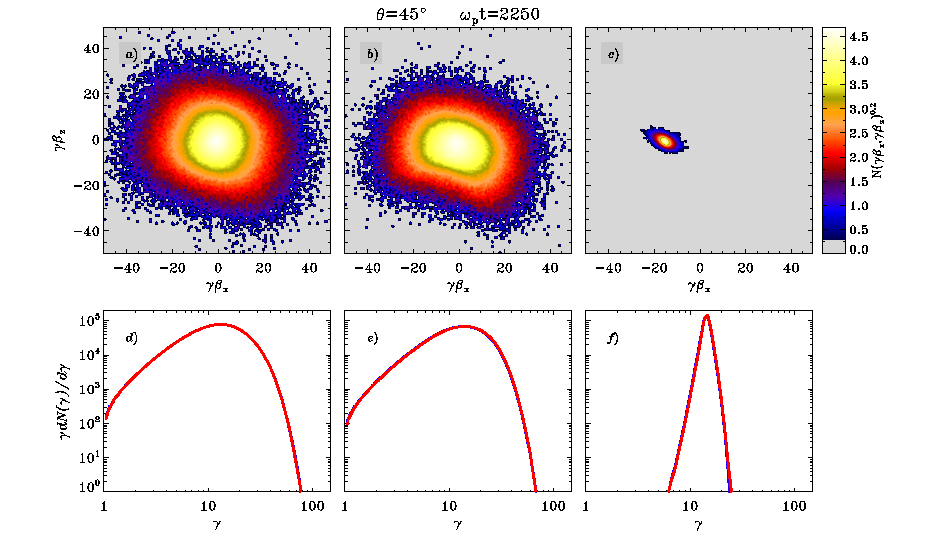}
\caption{Positron momentum space and electron (blue) and positron (red) spectra at time $\omega_{\rm{p}}t=2250$ for $\theta=45^\circ$, at three slices through the flow as marked by arrows at the bottom of Fig.~\fig{fluid45}. See the caption of Fig.~\fig{subspace0} for details.}
\label{fig:subspace45}
\end{center}
\end{figure*}

\subsection{Superluminal shocks: $\theta_{\rm crit}\lesssim\theta\leq90\deg$}\label{sec:super}
Figs.~\fig{fluid45} and \fig{subspace45} show the shock structure, phase space plots and particle spectra for a representative superluminal shock ($\theta=45^\circ$) at time $\ompt=2250$. No significant particle acceleration occurs in superluminal shocks, and the upstream plasma is not appreciably perturbed by the presence of the shock.

In the phase space plot of Fig.~\fig{fluid45}\textit{\,l}, the incoming plasma is seen as a cold beam with $\gamma\beta_{\rm x}\approx-15$, which is slightly heated while approaching the shock and then sharply thermalized in the shock transition layer at $x_{\rm sh}\approx920\comp$. As discussed by \citet{hoshino_91} and anticipated in \S\ref{sec:oblique}, the incoming plasma thermalizes due to the absorption of cyclotron waves produced by coherent particle gyration in the compressed shock fields. Thermalization proceeds with time in the downstream medium: panels \tit{l} and \tit{m} of Fig.~\fig{fluid45} show that the plasma farther downstream, which had more time to thermalize after passing through the shock, has a broader $x-\gb{x}$ and $x-\gb{y}$ distribution than the fluid closer to the shock. 

Downstream particles preferentially move in a plane perpendicular to the downstream oblique field; indeed, the oblique boxy shape of their downstream $\gb{x}-\gb{z}$ momentum space (Figs.~\fig{subspace45}\textit{\,a},\,\tit{b} for two different downstream slabs) is suggestive of particle gyro-motion around the oblique field. As discussed at the beginning of  \S\ref{sec:struct}, for magnetic obliquities $\theta\gtrsim45\deg$ it is harder for downstream particles to isotropize along the magnetic field rather than in the plane orthogonal to it. Since for high obliquities such plane approaches the $xy$ simulation plane, this should result in the downstream thermal width of the $x-\gb{z}$ phase space being smaller than the $x-\gb{x}$ or $x-\gb{y}$ distributions, as observed (compare panel \tit{n} with panels \tit{l} and \tit{m} in Fig.~\fig{fluid45}).

The net fluid velocity $\beta_{\rm z,d}<0$  behind the shock in Fig.~\fig{fluid45}\textit{\,n} is responsible for the residual motional electric field $E_{\rm y,d}>0$ observed in the simulation frame downstream from the shock (hard to see in Fig.~\fig{fluid45}\textit{\,h}, but see Table \ref{tab1}), a feature common to all oblique magnetic configurations, as discussed in \S\ref{sec:struct}. This residual fluid velocity vanishes farther downstream, in regions already overtaken by the slow shock. The slow shock appears as a smooth increase in density (Fig.~\fig{fluid45}\textit{\,b}) and decrease in magnetic energy (Fig.~\fig{fluid45}\textit{\,d}) and $B_{\rm z}$ (Fig.~\fig{fluid45}\textit{\,f}) behind the main shock towards the wall. Similar properties for the slow shock are observed for $\theta=30\deg$ in Fig.~\fig{fluid30}, whereas the slow shock for $\theta=15^\circ$ is outside the $x$-range shown in Fig.~\fig{fluid15}.

Most notably, no evidence for particle acceleration is present for $\theta=45\deg$. No high-energy particles are observed to propagate upstream from the shock (Fig.~\fig{fluid45}\textit{\,l-n}), and the particle energy spectrum in upstream slices is just the  thermal distribution of the injected plasma Lorentz-boosted along $-\bf{\hat{x}}$ (Fig.~\fig{subspace45}\textit{\,f}). The range of downstream 4-velocities is much smaller than for subluminal shocks (compare panels \textit{l-n} in Fig.~\fig{fluid45} with the same panels in Figs.~\fig{fluid0},\,\fig{fluid15},\,\fig{fluid30}), and the downstream energy spectrum (Fig.~\fig{subspace45}\textit{\,d},\,\tit{e}) resembles a purely thermal distribution without any suprathermal tail. The lack of  returning particles explains why no oblique waves are seen in the upstream medium of superluminal shocks (compare panels \textit{\,e},\,\tit{g},\,\tit{i} in Fig.~\fig{fluid45} with the same panels in Figs.~\fig{fluid0},\,\fig{fluid15},\,\fig{fluid30}).

We remark that the main features discussed here for $\theta=45^\circ$, and most importantly the absence of shock-accelerated particles, are common to all superluminal configurations up to the perpendicular case $\theta=90^\circ$ investigated via PIC simulations by, e.g., \citet{gallant_92} and \citet{spitkovsky_05}. 

\section{Particle energy spectra}\label{sec:spectra}
We now investigate particle acceleration in subluminal and superluminal shocks by studying the downstream energy spectra of our representative sample of magnetic obliquities: $\theta=0\deg$, $15\deg$, $30\deg$ (subluminal shocks) and $\theta=45\deg$ (superluminal shock). We remind that the boundary between subluminal and superluminal configurations is $\thetacrit\approx34\deg$.

In \S\ref{sec:dep_mag} we show that subluminal shocks ($0\leq\theta\lesssim\thetacrit$) produce a population of high-energy  particles as part of the shock evolution. Such shock-accelerated particles populate a suprathermal power-law tail in the energy spectrum behind the shock. As the magnetic obliquity increases within the subluminal range, the high-energy tail becomes flatter and it accounts for an increasing fraction of downstream particle number and energy. In other words, the acceleration efficiency increases with magnetic obliquity for $\theta\lesssim\thetacrit$. Instead, the downstream particle spectrum for superluminal shocks, $\thetacrit\lesssim\theta\leq90\deg$ (including perpendicular shocks), does not show any significant suprathermal tail, meaning that particle acceleration is extremely inefficient for superluminal configurations.

In \S\ref{sec:time} we follow the time evolution of downstream particle spectra and show that the slope and normalization of the high-energy tail in high-obliquity -- yet still subluminal -- shocks varies with time much more than for low obliquities. This suggests that a different acceleration mechanism is operating in shocks close to the superluminality threshold as compared to quasi-parallel shocks, as we discuss in \S\ref{sec:mechanism}.

In the following, all the particle spectra are computed in a slab $200\,c/\omega_{\rm p}$-wide which is centered $400\,c/\omega_{\rm p}$ downstream from the shock, regardless of time or magnetic obliquity. Strictly speaking, our particle spectra are not computed in the downstream fluid frame, due to the net fluid velocity $\beta_{\rm z,d}$ observed in the simulation frame behind oblique shocks; however, $|\beta_{\rm z,d}|\lesssim0.1$, so that  our wall-frame particle spectra are an excellent proxy for their downstream-frame counterparts. In order to estimate the acceleration efficiency of subluminal shocks, we fit the downstream particle spectrum with a three-dimensional Maxwellian  (as appropriate for  $\theta\lesssim45\deg$ shocks, see \S\ref{sec:struct}) with spread $\Delta\gamma_{\msc{mb}}$ plus a power law starting at $\gamma_{\rm min}$ with exponential cutoff at $\gamma_{\rm cut}$ whose width is $\Delta\gamma_{\rm cut}$. The fitting function is $f(\gamma)=N_{\msc{mb}}\gamma\sqrt{\gamma^2-1}\exp(-\gamma/\Delta\gamma_{\msc{mb}})+N_{\msc{pl}}\gamma^{-p}\rm{min}[1,\exp(-(\gamma-\gamma_{\rm cut})/\Delta\gamma_{\rm cut})]$, with $N_{\msc{pl}}=0$ for $\gamma<\gamma_{\rm min}$. The high-energy tail is defined to start at the intersection of the fitting Maxwellian and the fitting power law.

\begin{figure*}[htbp]
\begin{center}
\includegraphics{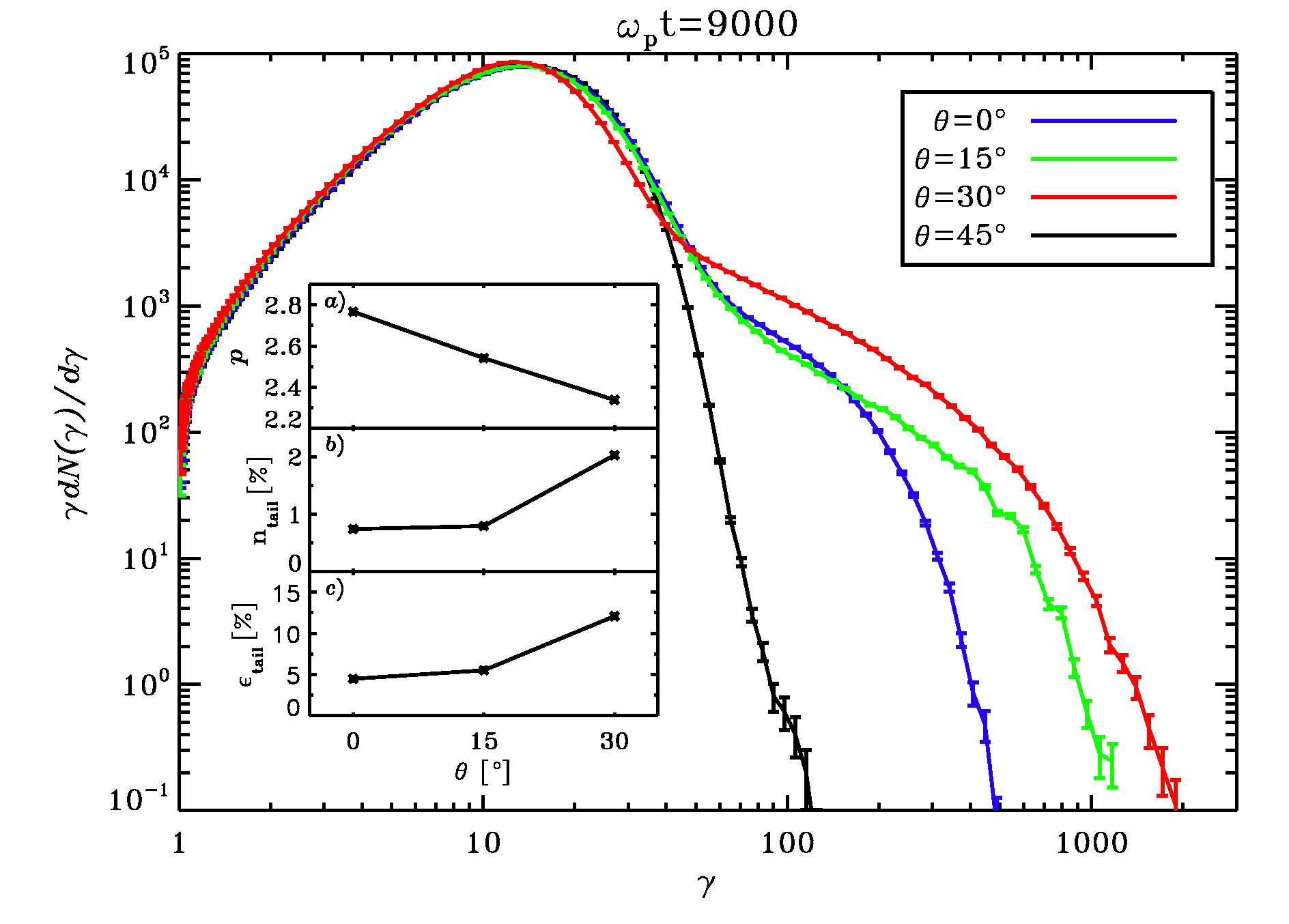}
\caption{Downstream particle energy spectra at time $\omega_{\rm p} t=9000$ for different magnetic obliquities: $\theta=0^\circ$ (blue), $15^\circ$ (green), $30^\circ$ (red) and $45^\circ$ (black). All spectra are computed in a slab $200\, c/\omega_{\rm p}$-wide centered $400\, c/\omega_{\rm p}$ downstream from the shock. The spectrum is normalized to the total number of particles within the considered slab, apart from a constant multiplication factor; it is supplemented with Poissonian error bars.  Subpanels: \tit{a}) Power-law slope of the high-energy tail as a function of the obliquity angle $\theta$. The tail is defined to start at the intersection between the fitting Maxwellian and the fitting power law. \tit{b}) Fraction of particles (in percent) in the high-energy tail. \tit{c}) Fraction of energy (in percent) stored in the high-energy tail.}
\label{fig:spectot}
\end{center}
\end{figure*}

\subsection{Dependence on magnetic obliquity}\label{sec:dep_mag}
Fig.~\fig{spectot} shows the main result of our work, namely the dependence of the acceleration efficiency on magnetic obliquity: downstream spectra for $\theta=0^\circ$ (blue), $15^\circ$ (green), $30^\circ$ (red) and $45^\circ$ (black) are shown at the final time $\omega_{\rm p}t=9000$ of our simulation runs. Subpanel \tit{a} shows the slope $p$ of the fitting power law as a function of magnetic obliquity, while subpanels \tit{b} and \tit{c} present how the fraction of particles and energy stored in the suprathermal tail depends on the magnetic inclination.

For subluminal shocks, the downstream energy spectrum consists of a low-energy thermal bump and a high-energy suprathermal tail. The low-energy part, which accounts for most  of the energy, is populated by the particles that are directly transmitted downstream, where they are isotropized and thermalized. A fraction of the incoming particles may be scattered or reflected back upstream when they first encounter the shock; they remain in the vicinity of the shock for longer times and may be significantly accelerated (see the high-energy population at the shock in panels \tit{l-n} of Figs.~\fig{fluid0},\,\fig{fluid15},\,\fig{fluid30}). If they finally escape upstream, they may contribute to the beam of returning particles which trigger the generation of the upstream oblique waves; if transmitted downstream, they populate the high-energy suprathermal tail of downstream particle spectra. 

From Fig.~\fig{spectot} it is apparent that the suprathermal tail  in subluminal shocks (blue, green and red lines for $\theta=0\deg$, $15\deg$ and $30\deg$ respectively) is much more pronounced than in superluminal shocks (black line for $\theta=45\deg$), whose particle spectrum does not significantly deviate from a thermal distribution. The suprathermal tail becomes flatter and more populated as the magnetic obliquity increases within the subluminal range, meaning that shocks close to the superluminality threshold $\thetacrit\approx34\deg$ accelerate more efficiently  than quasi-parallel shocks.  As  $\theta$ increases from $0^\circ$ to $30^\circ$, the slope of the power-law tail  increases from $-2.8\pm0.1$ to $-2.3\pm0.1$ (subpanel \textit{a}), the fraction of particles in the tail grows from $\sim1\%$ to $\sim2\%$ (subpanel \textit{b}) and the energy fraction they account for increases from $\sim4\%$ to $\sim12\%$ (subpanel \textit{c}); at the end of our simulations the spectrum extends up to $\gamma_{\rm max}\approx500$ for $\theta=0\deg$ and reaches $\gamma_{\rm max}\approx2000$ for $\theta=30\deg$. The negligible suprathermal tail observed for $\theta=45\deg$ is much steeper than for subluminal shocks; as a conservative upper limit, the fraction of particles and energy it accounts for is smaller by two orders of magnitude than in subluminal configurations.   

The dependence of the acceleration efficiency on magnetic obliquity should affect the temperature $\Delta\gamma_{\msc{mb}}$ of the low-energy Maxwellian; we expect that, the larger the fraction of energy stored in the suprathermal tail, the cooler the low-energy bump, if the mean downstream particle Lorentz factor $\bar\gamma$ were constant with magnetic obliquity. However, with increasing $\theta$, a larger fraction of the upstream kinetic energy is converted to downstream magnetic energy (see Table \ref{tab1}), so that the average downstream particle energy monotonically decreases with $\theta$. In summary, we expect that, within the subluminal range, $\Delta\gamma_{\msc{mb}}$ should systematically decrease with $\theta$, both because $\bar\gamma$ decreases and because a larger fraction of downstream particle energy is stored in the suprathermal tail; across the boundary between subluminal and superluminal configurations, $\Delta\gamma_{\msc{mb}}$ should sharply increase, if the drop in acceleration efficiency prevails over the decrease of $\bar\gamma$ with $\theta$. Indeed, the thermal spread of the fitting Maxwellian decreases from $\sim4.6$ for $\theta=0\deg$ to $\sim4.2$ for $\theta=30\deg$, and then jumps back to $\sim4.5$ for $\theta=45\deg$; in Fig.~\fig{spectot}, such trend can be seen in the position of the peak of the low-energy bump. 

In summary, the high-energy tail becomes harder and more populated as the magnetic obliquity increases within the subluminal range, in agreement with semi-analytical \citep[e.g.,][]{kirk_heavens_89, ballard_heavens_91} and Monte-Carlo \citep[e.g.,][]{ostrowski_bednarz_98, niemiec_ostrowski_04,ellison_double_04} calculations. For superluminal shocks, we find that self-generated turbulence is not sufficiently strong to trap a significant fraction of particles in the vicinity of the shock for efficient acceleration. As we show in \S\ref{sec:mechanism}, the upstream oblique waves discussed in \S\ref{sec:sub} regulate  injection into the acceleration process for $\theta\lesssim\theta_{\rm crit}$. Since no waves are present in the upstream medium of superluminal shocks, due to the absence of the stream of returning particles that trigger their growth, a negligible fraction of the incoming particles enters the acceleration process, and the acceleration efficiency is correspondingly small. The suprathermal tail barely seen for $\theta=45^\circ$ in Fig.~\fig{spectot} is probably populated by particles accelerated during a single shock crossing from upstream to downstream, and then advected downstream. We further explore the transition between subluminal and superluminal shocks with regards to downstream particle spectra in Appendix B, where we show that the drop in acceleration efficiency  across $\thetacrit\approx34\deg$ is rather abrupt.

Finally, we remind that in Fig.~\fig{spectot} we are comparing downstream spectra computed in a slab whose distance behind the shock is the same for all magnetic obliquities. This may be misleading, since the acceleration process in subluminal shocks is not in steady state yet (see \S\ref{sec:time}), and downstream particle spectra at a given time may vary among different longitudinal locations (e.g., compare panels \tit{d} and \tit{e} in Fig.~\fig{subspace30}). The particle energy spectrum in a slab at fixed distance downstream from the shock is affected by the efficiency both in accelerating particles and in transmitting downstream the freshly-accelerated particles. Since the latter decreases with increasing obliquity if cross-field diffusion is of minor importance, shocks of larger obliquities may appear, as evaluated from a slab at fixed distance behind the shock, to accelerate less efficiently just because particle diffusion downstream from such shocks is slower. By comparing particle spectra in a slab at constant post-shock distance (Fig.~\fig{spectot}), we find that subluminal shocks with higher obliquities accelerate more efficiently; this will hold true \emph{a fortiori} when taking into account the obliquity dependence of the downstream diffusion probability.

\begin{figure}[tbp]
\begin{center}
\includegraphics[width=0.5\textwidth]{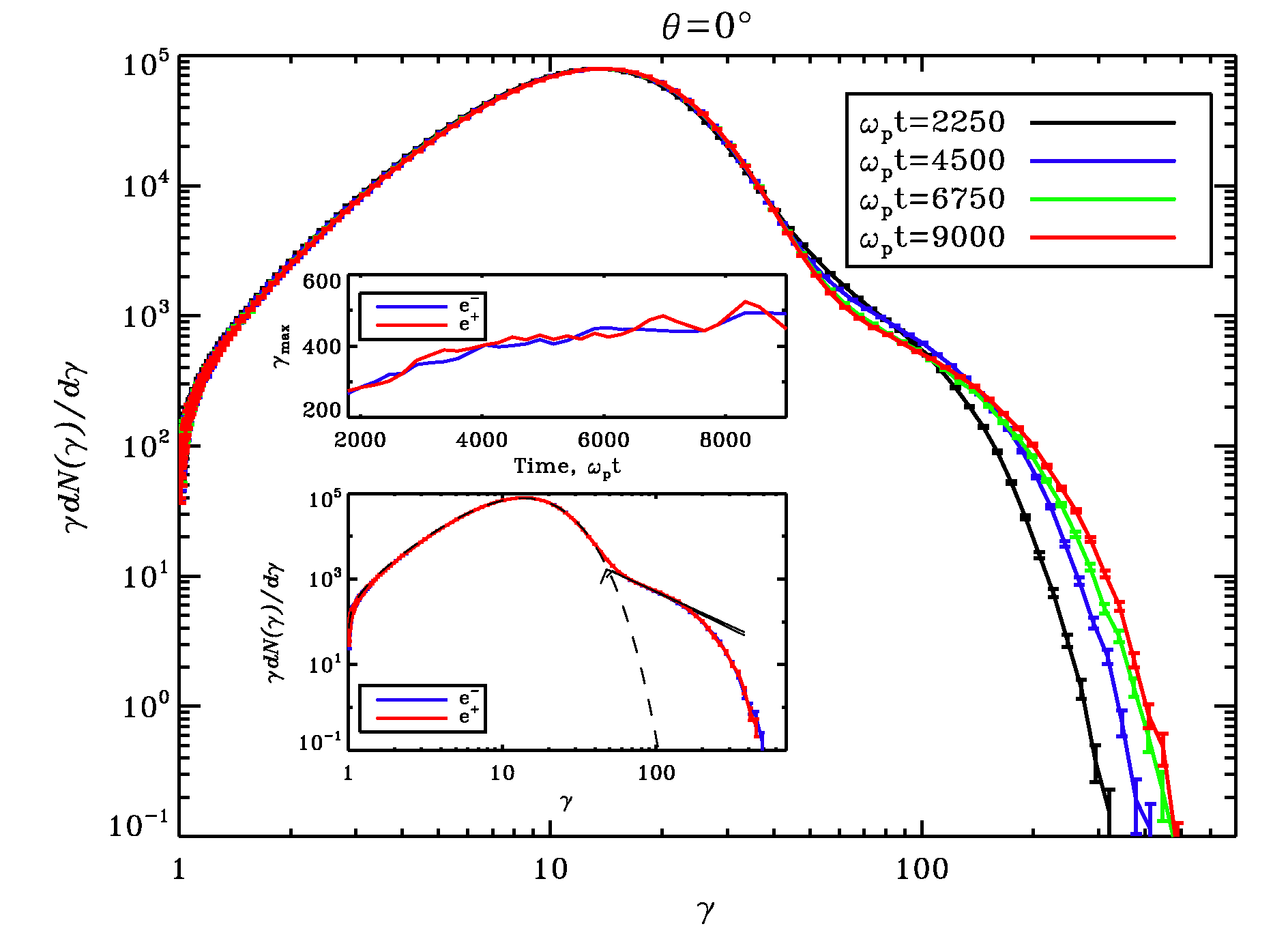}
\caption{Time evolution of the particle energy spectrum for $\theta=0^\circ$: $\omega_{\rm{p}}t=2250$ (black), $\omega_{\rm{p}}t=4500$ (blue), $\omega_{\rm{p}}t=6750$ (green), $\omega_{\rm{p}}t=9000$ (red). All spectra are computed in a slab $200\, c/\omega_{\rm p}$-wide centered $400\, c/\omega_{\rm p}$ downstream from the shock. Upper subpanel: electron (blue) and positron (red) maximum Lorentz factor within the slab where the spectrum is computed, as a function of time. Lower subpanel: electron (blue) and positron (red) spectra at time $\omega_{\rm{p}}t=9000$ with Poissonian error bars, supplemented by the best-fitting Maxwellians (black dashed lines) and power laws (black solid lines).}
\label{fig:spectime0}
\end{center}
\end{figure}

\begin{figure}[tbp]
\begin{center}
\includegraphics[width=0.5\textwidth]{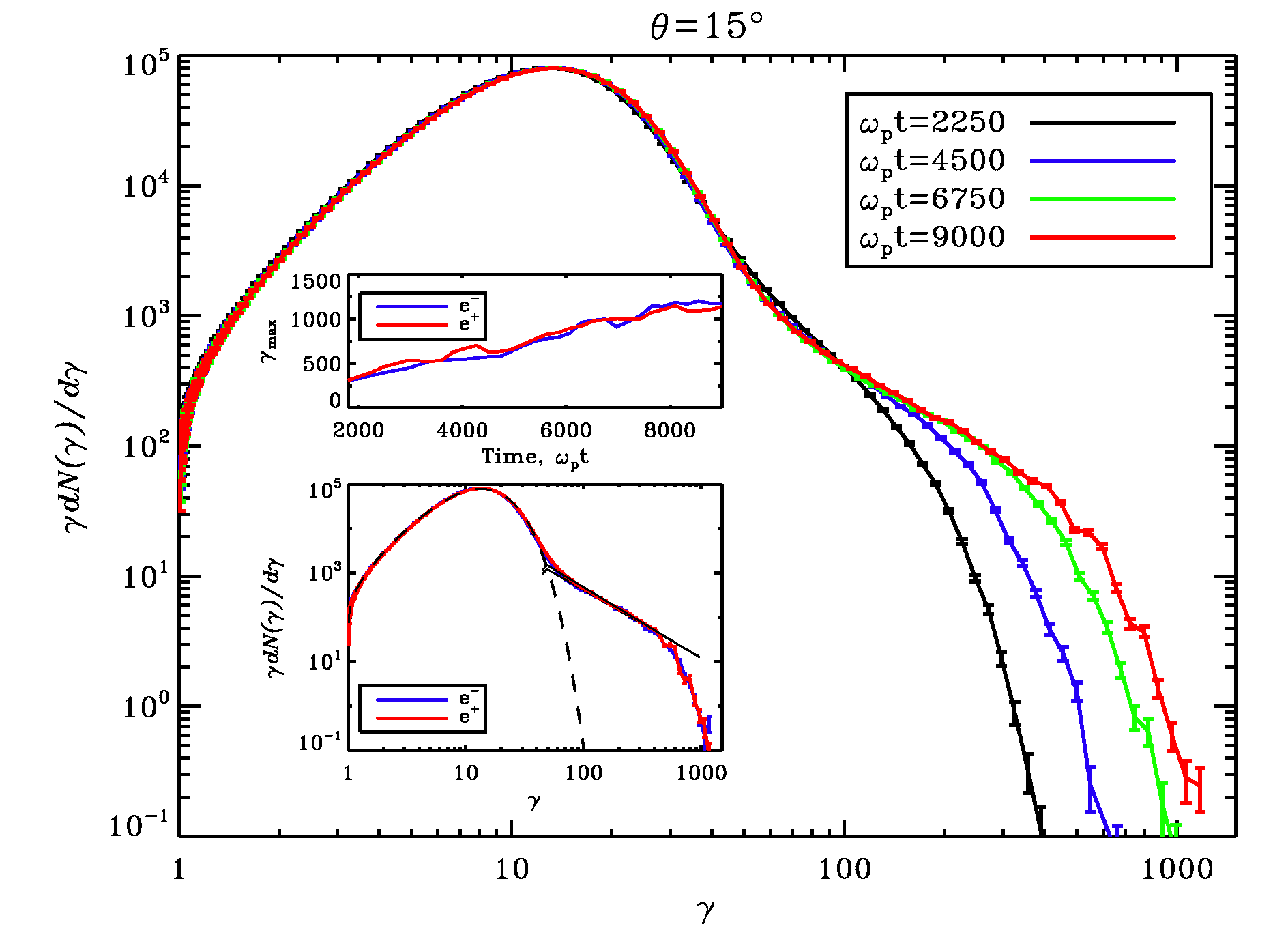}
\caption{Time evolution of the particle energy spectrum for $\theta=15^\circ$. See the caption of Fig.~\fig{spectime0} for details.}
\label{fig:spectime15}
\end{center}
\end{figure}

\begin{figure}[tbp]
\begin{center}
\includegraphics[width=0.5\textwidth]{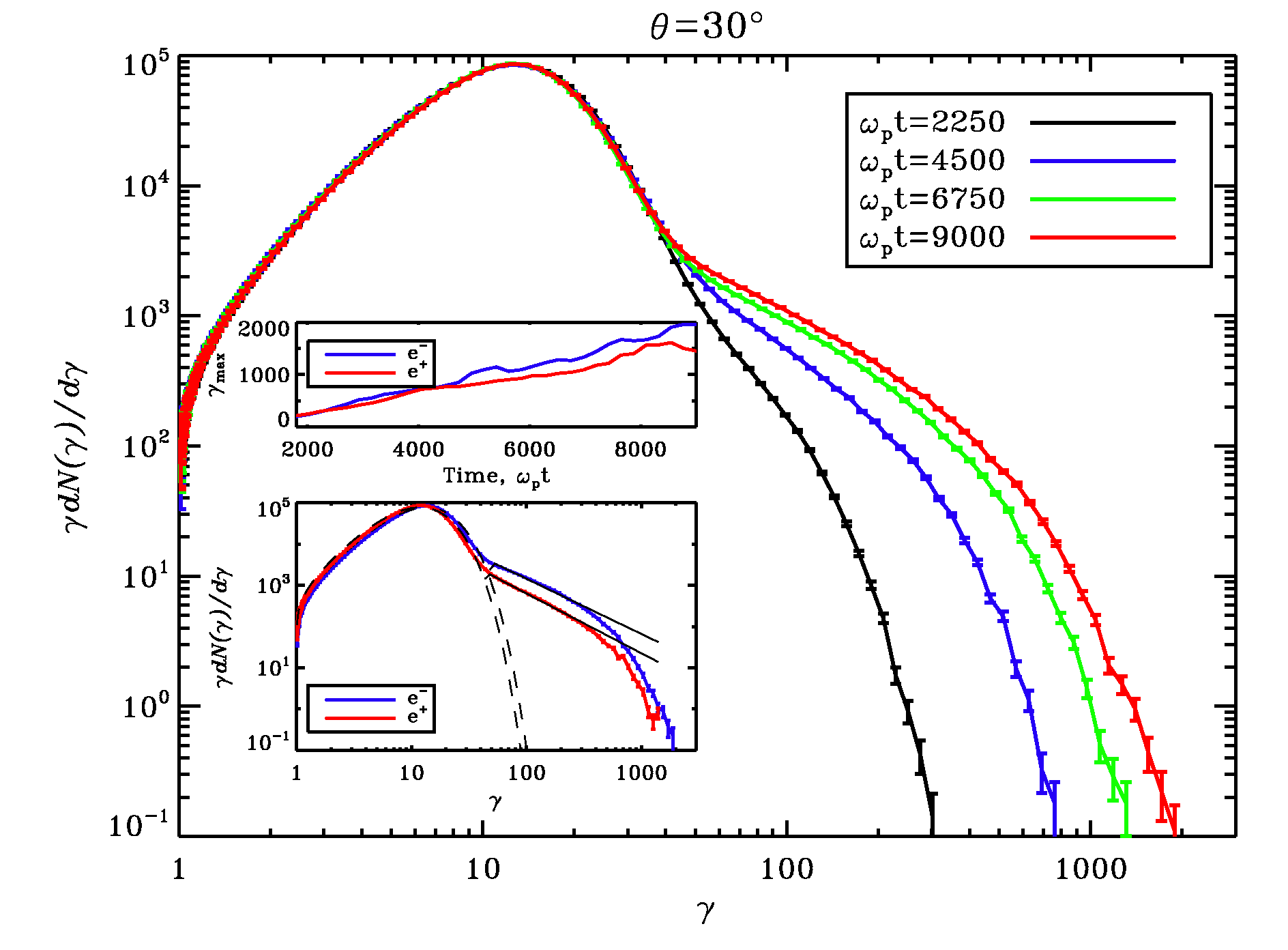}
\caption{Time evolution of the particle energy spectrum for $\theta=30^\circ$. See the caption of Fig.~\fig{spectime0} for details.}
\label{fig:spectime30}
\end{center}
\end{figure}

\begin{figure}[tbp]
\begin{center}
\includegraphics[width=0.5\textwidth]{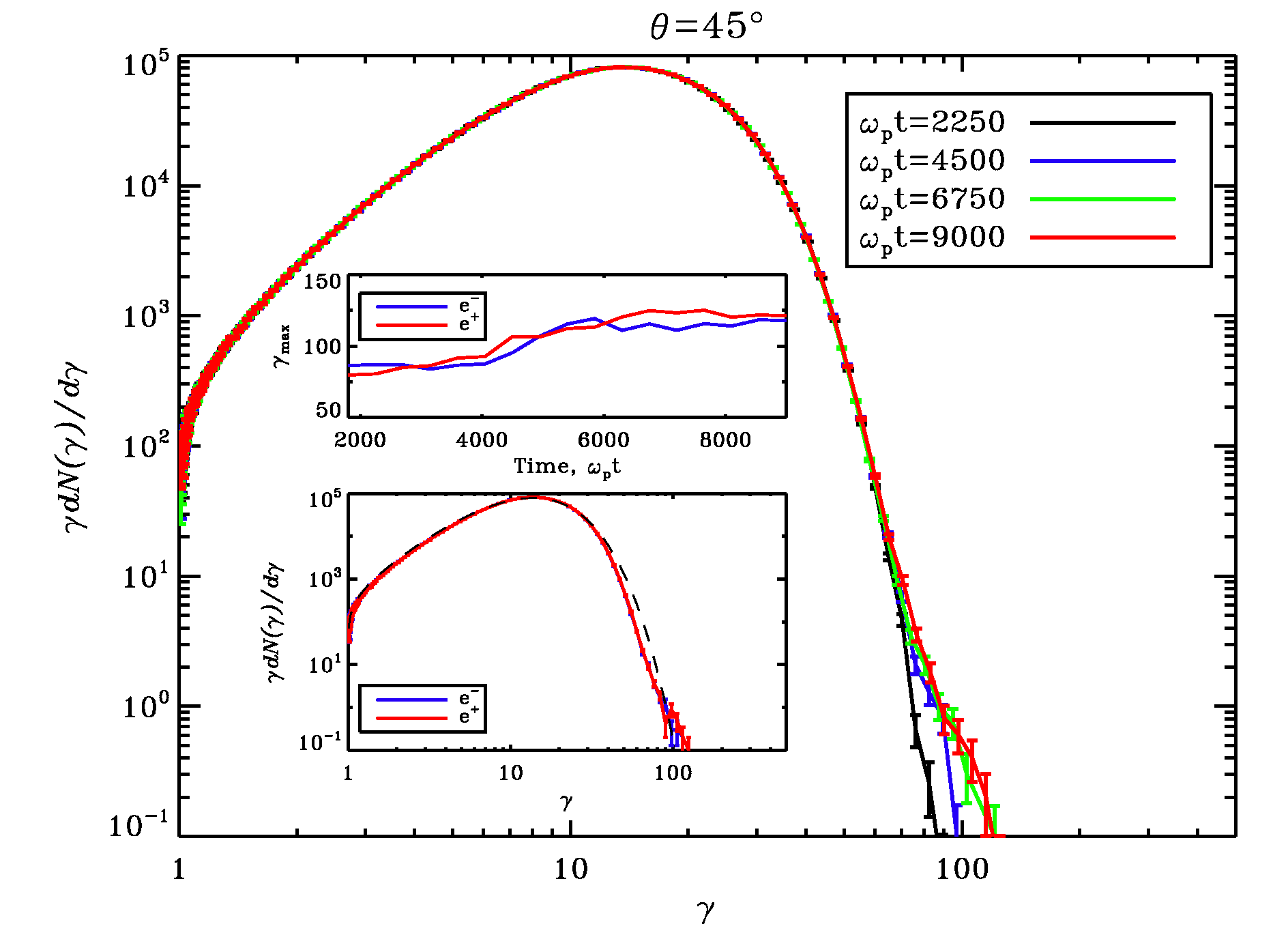}
\caption{Time evolution of the particle energy spectrum for $\theta=45^\circ$. See the caption of Fig.~\fig{spectime0} for details. Here, we do not attempt to fit the suprathermal tail.}
\label{fig:spectime45}
\end{center}
\end{figure}

\subsection{Time evolution}\label{sec:time}
Figs.~\fig{spectime0},\,\fig{spectime15} and \fig{spectime30} follow the time evolution of downstream particle spectra for our representative sample of subluminal shocks ($\theta=0\deg$, $15\deg$ and $30\deg$ respectively), whereas Fig.~\fig{spectime45} refers to the superluminal shock $\theta=45\deg$. In the upper subpanel of each figure, we plot the time evolution of the electron (blue) and positron (red) maximum Lorentz factor in the downstream slab where the spectrum is computed. The lower subpanel shows the electron (blue) and positron (red) spectra at the final time of our simulations ($\ompt=9000$), together with the best-fitting low-energy Maxwellians (black dashed lines) and high-energy power laws (black solid lines).

For both subluminal and superluminal magnetic configurations, the low-energy thermal bump  does not significantly evolve with time. Instead, the high-energy tail grows with time as more and more particles are shock-accelerated, but its evolution is much more significant for subluminal shocks than for superluminal configurations.  For $\theta=45\deg$ (Fig.~\fig{spectime45}), the particle spectrum for early times (black line) is remarkably thermal, and the minor suprathermal tail visible at later times stops evolving after $\ompt=6650$ (compare green and red lines in Fig.~\fig{spectime45}). The upper subpanel in Fig.~\fig{spectime45} confirms that the maximum particle Lorentz factor saturates at $\gamma_{\rm max}\approx120$ after $\ompt\approx5000$. We are then reasonably confident that  for $\theta=45\deg$ our simulation captures the full time evolution of the downstream particle spectrum, so that the claimed inefficiency of particle acceleration in superluminal shocks should still hold true with a longer simulation timespan.

For subluminal shocks, the high-energy tail stretches and flattens with time. Its time evolution  is moderate for low obliquities ($\theta=0\deg$ in Fig.~\fig{spectime0} and $\theta=15\deg$ in Fig.~\fig{spectime15}) and more dramatic for magnetic inclinations close to the superluminality threshold ($\theta=30\deg$ in Fig.~\fig{spectime30}).  For $\theta=0\deg$ the maximum particle Lorentz factor   increases  from $\gamma_{\rm max}\approx300$ at $\ompt=2250$ to $\gamma_{\rm max}\approx500$ at $\ompt=9000$, whereas for $\theta=30\deg$ it grows much faster, from $\gamma_{\rm max}\approx300$ to $\gamma_{\rm max}\approx2000$. The fraction of particles and energy stored in the tail also increases during our simulations. Therefore, for subluminal obliquities (and especially for angles $\theta\lesssim\thetacrit$, which show the most significant time evolution), the shock acceleration process is not in steady state yet, and the values we quote for the fractional contribution by the suprathermal tail to the downstream particle and energy census should be taken as lower limits. On the other hand, for late times (compare the green and red lines in Figs.~\fig{spectime0},\,\fig{spectime15},\,\fig{spectime30}), the shape of the spectrum does not change significantly, apart from the linear increase with time of the upper energy cutoff. Since the power-law spectral index is always $p>2$, this means that the values we quote for the acceleration efficiency at $\ompt=9000$ will not significantly increase at later times.

Concerning the time evolution of the maximum particle Lorentz factor, the upper subpanels of Figs.~\fig{spectime15} and \fig{spectime30} (for $\theta=15\deg$ and $30\deg$ respectively) show a roughly linear increase with time, whereas for $\theta=0\deg$ it saturates at $\gamma_{\rm max}\approx500$ after $\ompt\approx6000$ (upper subpanel in Fig.~\fig{spectime0}). By running simulations of parallel shocks in which the first backward jump of the injector happens at different times (we tried 30000, 40000 and 50000 timesteps), we observe that, the sooner the injector starts jumping, the quicker the acceleration process saturates. Each jump unphysically removes the returning particles which are farthest upstream. This has two consequences: first, we artificially prevent these particles from returning to the shock and possibly contributing to the downstream spectrum; also, since the stream of returning particles triggers the growth of the upstream oblique waves, which play a major role in the acceleration process (see \S\ref{sec:mechanism}),  we are unphysically affecting the whole process of shock-acceleration. In summary, the spectral time-evolution for $\theta=0\deg$ saturates due to our jumping injector, and the acceleration efficiency and maximum Lorentz factor we report for $\theta=0^\circ$ should be taken as lower limits. Instead, for higher obliquities ($\theta=15\deg$ and $30\deg$), no unphysical saturation is observed in the spectrum, since the returning particles remain closer to the shock, so that fewer particles are removed by our jumping injector. The jumping injector does not seem to cause other unphysical effects on particle acceleration. 

Finally, we point out that the difference between the downstream spectra of positrons and electrons for $\theta=30\deg$ (lower subpanel of Fig.~\fig{spectime30}, at $\ompt=9000$), which is reminescent of an analogous difference observed at earlier times (Fig.~\fig{subspace30}\pan{e}, at $\ompt=2250$), is an artifact of the small transverse size of our simulations.  For magnetic obliquities near  $\theta=30\deg$,  the stream of returning particles, which triggers the growth of the upstream oblique waves, does not propagate very far  from the shock. As a consequence, the upstream pattern of electrostatic waves does not move far enough from the shock to let the shock electric potential fluctuate in time (Fig.~\fig{fluid30}\textit{\,k} for $\ompt=2250$). Particles of one charge will then enter the shock with  energies systematically higher than particles of the opposite charge, and the spectral difference between electrons and positrons will persist over time. For smaller obliquities, the stream of returning particles recedes from the shock at higher speeds and the time-average of the shock electric potential vanishes, so that no systematic difference is observed between the downstream spectra of electrons and positrons for $\theta=0\deg$ and $\theta=15\deg$ (lower subpanels in Figs.~\fig{spectime0} and \fig{spectime15}). 
For $\theta=30\deg$, we have tested that the downstream spectral difference disappears when the simulation box becomes larger (more than 4096 transverse cells) than the characteristic transverse coherence length of the upstream electrostatic oscillations. Therefore, no spectral asymmetry between electrons and positrons is expected in any realistic astrophysical scenario. Moreover, the sum of electron and positron spectra does not vary among runs with different box width, proving that the shock electric potential does not play any appreciable role in the overall acceleration process.\footnote{As we comment in \S\ref{sec:disc}, this is no longer true for electron-ion shocks, where ion acceleration via the so called Shock-Surfing Acceleration mechanism depends on the shock electric potential.}

\section{Particle acceleration mechanisms}\label{sec:mechanism}
We explore the mechanisms responsible for particle acceleration in subluminal shocks by tracing the trajectories of representative high-energy particles extracted from the simulations.  We find that particle acceleration is mostly mediated by Diffusive Shock Acceleration (DSA) for quasi-parallel shocks ($\theta\lesssim10^\circ$), but Shock-Drift Acceleration (SDA) is the main acceleration mechanism for larger, yet still subluminal, magnetic obliquities. 

\begin{figure*}[htbp]
\begin{center}
\includegraphics{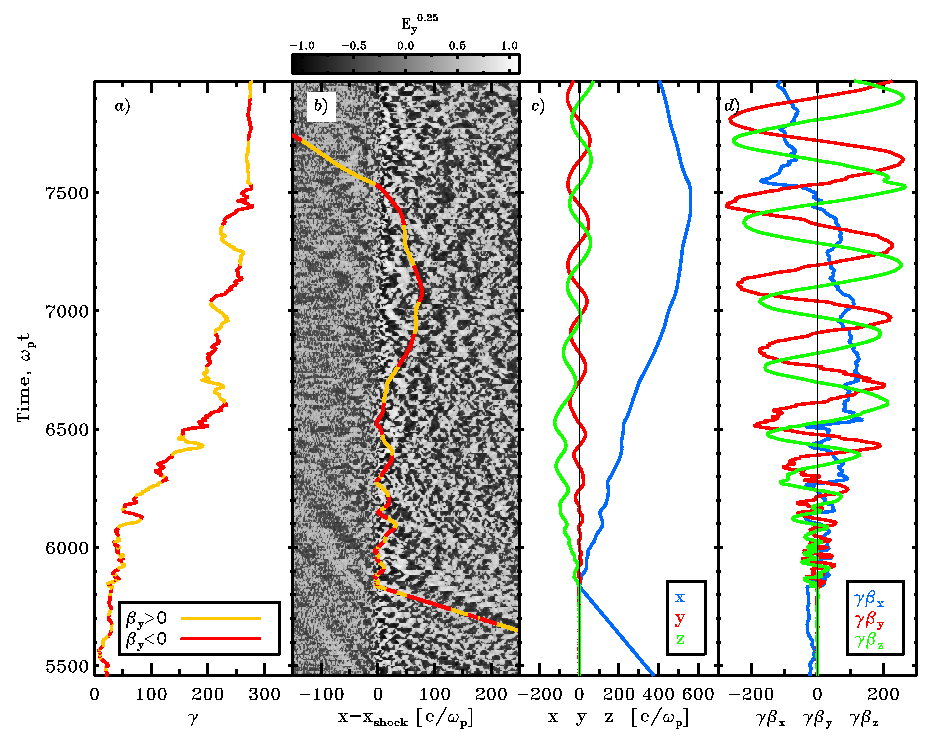}
\caption{Trajectory of a representative high-energy positron for $\theta=0^\circ$, in the wall frame of the simulation: \tit{a}) Particle Lorentz factor as a function of time; colors show when the particle $y$-velocity is positive (yellow) or negative (red). \tit{b}) Longitudinal position relative to the shock as a function of time, with the same color coding as in panel \textit{a}. The particle orbit is superimposed over two-dimensional strips of $E_{\rm y}$ stacked in time.   For every piece of the particle trajectory that lasts $22.5\,\omega_{\rm p}^{-1}$, we plot a $5\comp$-wide $E_{\rm y}$ strip which is centered along $y$ at the particle transverse location after $11.25\,\omega_{\rm p}^{-1}$ since the start of that portion of the trajectory. Each strip extends along the whole $x$ dimension. $E_{\rm y}^\alpha$ stands for $(E_{\rm y}/|E_{\rm y}|)\,|E_{\rm y}|^\alpha$. $E_{\rm y}$ is measured in the simulation frame, but it is shifted in space so that the shock appears stationary. \tit{c}) Particle orbit as a function of time ($x(t)$ blue, $y(t)$ red and $z(t)$ green), relative to the particle position at its first encounter with the shock. \tit{d}) Particle dimensionless 4-velocities $\gamma\beta_{\rm x}(t)$ (blue), $\gamma\beta_{\rm y}(t)$ (red) and $\gamma\beta_{\rm z}(t)$ (green) as functions of time.}
\label{fig:part0}
\end{center}
\end{figure*}

\begin{figure*}[htbp]
\begin{center}
\includegraphics{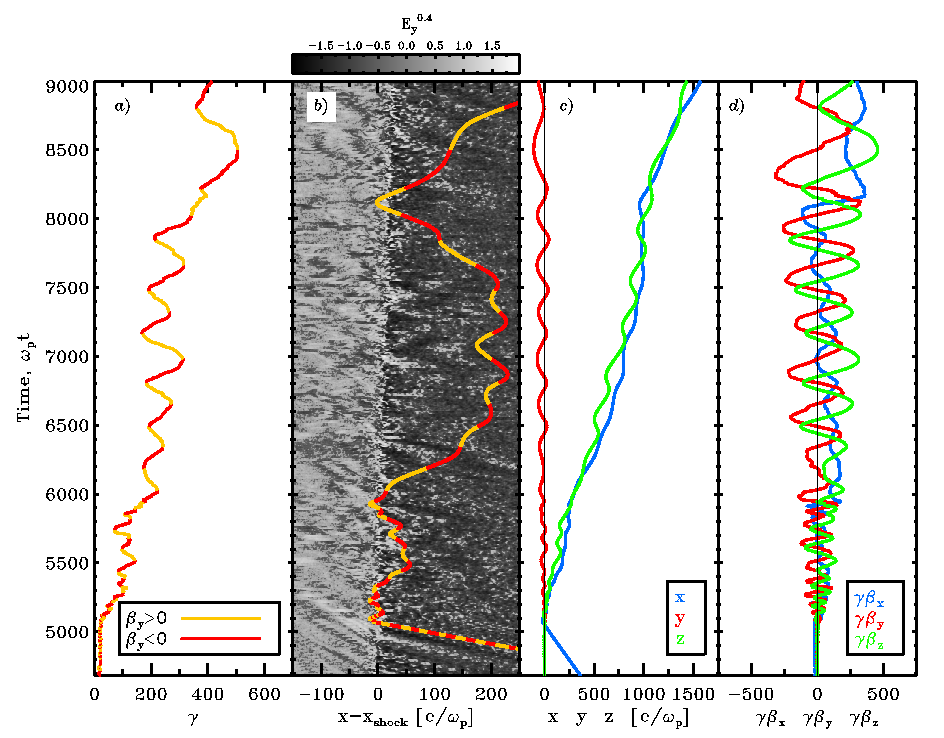}
\caption{Trajectory of a representative high-energy positron for $\theta=15^\circ$. See the caption of Fig.~\fig{part0}  for details.}
\label{fig:part15}
\end{center}
\end{figure*}

\begin{figure*}[htbp]
\begin{center}
\includegraphics{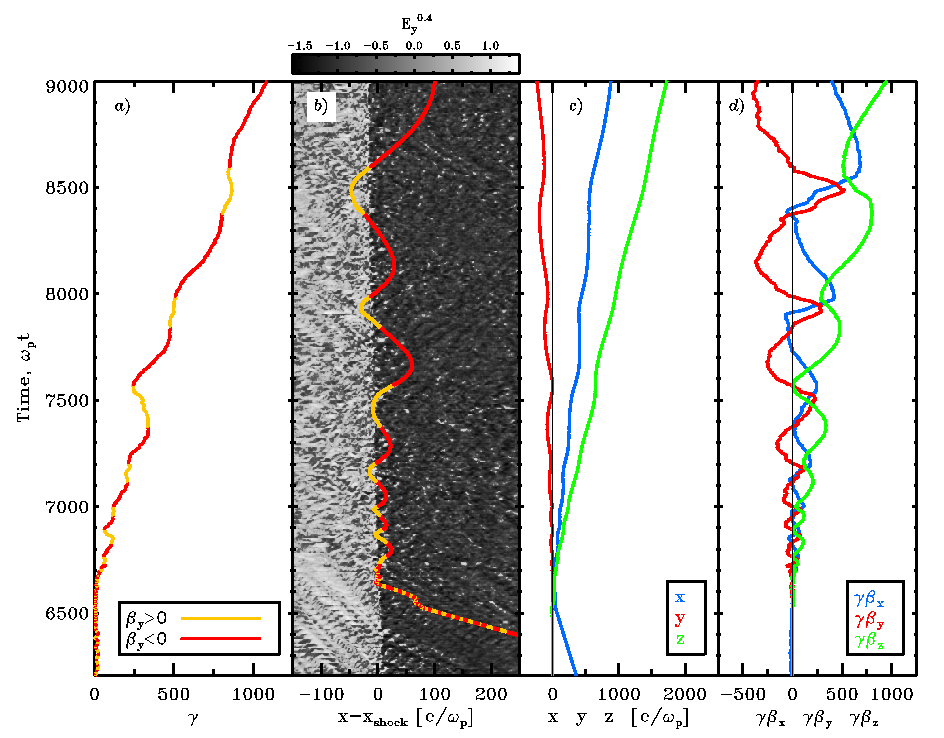}
\caption{Trajectory of a representative  high-energy positron for $\theta=30^\circ$. See the caption of Fig.~\fig{part0}  for details.}
\label{fig:part30}
\end{center}
\end{figure*}

The trajectories shown in Figs.~\fig{part0},\,\fig{part15} and \fig{part30} (for $\theta=0\deg$,\,$15\deg$ and $30\deg$ respectively) are characteristic of positrons that end up with the highest energies. In each figure, we show the time evolution of the particle Lorentz factor (panel \textit{a}) and 4-velocities (panel \textit{d}) in the simulation frame. In panel \tit{a}, red stands for negative $y$-velocity of the selected positron, yellow for positive $y$-velocity. Particle positions in the wall frame as a function of time are shown in panel \textit{c}. To clarify the interaction between particles and electromagnetic fields, in panel \textit{b} we plot the particle $x$-position relative to the shock (with the same color coding as in panel \tit{a}), superimposed over two-dimensional strips of $E_{\rm y}$ stacked in time. For every piece of the particle trajectory that lasts $22.5\,\omega_{\rm p}^{-1}$, we plot a $5\comp$-wide $E_{\rm y}$ strip which is centered along $y$ at the particle transverse location after $11.25\,\omega_{\rm p}^{-1}$ since the start of that portion of the trajectory. Each strip extends along the whole $x$ dimension.
The electric field is still measured in the wall frame but shifted along $x$ so that the shock appears stationary. We choose to plot $E_{\rm y}$ because it is the dominant electric component of the upstream oblique modes, and therefore it should be the primary agent of particle energization even for a parallel shock, where no uniform background $E_{\rm y,u} $ is present. 

In order to estimate the relative contribution of DSA and SDA to the acceleration process in oblique shocks ($\theta\neq0\deg$), we make use of the fact that particles are shock-drift accelerated while they drift parallel (or anti-parallel, according to their charge) to the background motional electric field. Since in the wall frame of our simulations the downstream motional electric field $E_{\rm y,d}$ is much smaller than the upstream $E_{\rm y,u}$ (see last column in Table \ref{tab1}), we may assume that  in the downstream medium  the particle energy stays approximately constant and we can relate the energy gain $\Delta\gamma_{\msc{sda}}$ due to SDA to the upstream drift $\Delta y_{\rm u}$ experienced by the particle:
\begin{equation}\label{eq:drift}
\Delta\gamma_{\msc{sda}}\sim\frac{q}{mc^2}E_{\rm y,u}\,\Delta y_{\rm u}
\end{equation}
where $q$ is the particle charge and $E_{\rm y,u}=-\beta_0B_{\rm z,u}<0$ the upstream background electric field. Since the right hand side of \eq{drift} can be computed from particle orbits, as well as the overall energy gain $\Delta\gamma$, the SDA fractional contribution $\Delta\gamma_{\msc{sda}}/\Delta\gamma$ will be a diagnostic of the relative importance of DSA and SDA for the acceleration process. In the extreme limit of scatter-free SDA we expect $\Delta\gamma_{\msc{sda}}/\Delta\gamma\sim1$.

\subsection{$\theta=0^\circ$}\label{sec:part0}
The orbit of a representative high-energy positron from the simulation of a parallel shock is plotted in Fig.~\fig{part0}. For $\omega_{\rm p}t\lesssim5800$, the particle is approaching the shock from upstream with velocity $\bmath{\beta}(t)\approx-\beta_0\,\hat{\mathbf{x}}$ (see $x(t)$, the blue line in panel \textit{c}). Since $y(t)$ stays approximately constant before the particle encounters the shock, the $E_{\rm y}$ strips stacked in panel \textit{b}   are all centered at the same $y$-location for $\ompt\lesssim5800$. Therefore, panel \tit{b} shows for $\ompt\lesssim5800$ how the $E_{\rm y}$ component of the upstream oblique modes (Fig.~\fig{fluid0}\textit{\,g} for $\ompt=2250$) evolves in time within a fixed longitudinal slice. It is thus evident that such waves are moving towards the shock and their phase velocity in the wall frame is smaller than the speed $\beta_0\approx1$ of the incoming particles. The positron energy varies with time even before it encounters the shock (Fig.~\fig{part0}\textit{\,a}), in response to oscillations in $\gamma\beta_{\rm x}(t)$ (blue line in Fig.~\fig{part0}\textit{\,d}) induced by the electrostatic component of the upstream waves.

From $\omega_{\rm p}t\approx5800$ to $\omega_{\rm p}t\approx6600$, the selected positron scatters between the shock and the upstream medium, and its energy is rapidly boosted up to a Lorentz factor $\gamma\approx250$ (Fig.~\fig{part0}\textit{\,a}). As the particle energy grows, its gyro-radius also proportionally increases, as seen in panel \textit{c} for $y(t)$ and $z(t)$. A moderate overall energy gain occurs upstream from the shock between $\omega_{\rm p}t\approx6600$ and $\omega_{\rm p}t\approx7500$, and finally at $\omega_{\rm p}t\approx7500$ the positron is transmitted downstream, where its energy remains constant. Most of the accelerated particles end up downstream, but the small fraction  that escapes upstream populates the returning stream which triggers the formation of the upstream oblique waves.

As seen in Fig.~\fig{part0}\textit{\,b}, the acceleration process involves stochastic bouncings between the shock and the upstream medium; magnetic turbulence in the shock transition layer prevents the particle from being transmitted downstream, whereas the upstream scattering is provided by the oblique waves discussed in \S\ref{sec:sub}, whose amplitude is largest close to the shock. In particular, Fig.~\fig{part0}\textit{\,b} suggests that the positron gains energy whenever its $y$-velocity $\beta_{\rm y}(t)$ (red if negative, yellow if positive) has the same sign as the electromagnetic field $E_{\rm y}$ (black if negative, white if positive), and decelerates if they have opposite signs. The converse will be true for electrons (not shown). Overall boosting of the selected positron results from favorable $\beta_{\rm y}(t) E_{\rm y}>0$ scatterings being more frequent than encounters in which $\beta_{\rm y}(t) E_{\rm y}<0$.

The stochastic nature of such wave-particle interactions, which are ultimately responsible for particle acceleration, is a defining feature of DSA. So, particle energization in parallel shocks proceeds via DSA, at the very least because the absence of a uniform background motional electric field rules out SDA. However, in the traditional picture of first-order Fermi acceleration, accelerated particles scatter off magnetic turbulence \textit{embedded} in the \textit{upstream} and \textit{downstream} media. Here, instead, high-energy particles bounce between the \textit{upstream} medium and the \textit{shock front}; moreover, the oblique modes responsible for the upstream scattering are not convected with the upstream plasma, as discussed above. Nevertheless, the stochastic character of the acceleration process points towards  a diffusive Fermi-like acceleration mechanism. Indeed, as theoretically predicted for diffusive acceleration in relativistic magnetized shocks \citep[e.g.,][]{achterberg_01}, we find that each acceleration cycle between the shock and the upstream medium involves an energy gain $\Delta \gamma\sim\gamma$, at least for the limited number of shock crossings ($\lesssim10$) observed in our simulation.

These results for a parallel shock should also apply to quasi-parallel magnetic configurations, in which the obliquity is small enough that the upstream motional electric field does not significantly affect the acceleration process. However, in \S\ref{sec:part15} we  show that SDA provides most of the energization for obliquity angles as low as $\theta=15^\circ$, so that DSA is the dominant acceleration mechanism only for $\theta\lesssim10\deg$.

\subsection{$\theta=15^\circ$}\label{sec:part15}
Fig.~\fig{part15} presents the acceleration process for a representative positron in a shock with $\theta=15\deg$. The particle is reflected back upstream on its first interaction with the shock ($\omega_{\rm p}t\approx5100$) and then gains energy at each shock encounter ($\omega_{\rm p}t\approx5200$, $\omega_{\rm p}t\approx5900$ and $\omega_{\rm p}t\approx8100$). The selected positron eventually escapes upstream at time $\omega_{\rm p}t\approx8200$ with Lorentz factor $\gamma\approx400$, contributing to the beam of returning high-energy particles. However, due to the limited timespan of our simulation, we cannot exclude the possibility that the positron will finally be deflected back and advected downstream.

The time evolution of the positron Lorentz factor plotted in panel \textit{a} suggests that energy gain is always associated with $\beta_{\rm y}(t)<0$ (red), i.e., when the positron $y$-velocity has the same sign as the uniform background electric field $E_{\rm y,u}<0$. Whenever the positron gyro-orbit is fully contained in the upstream region (e.g., from $\omega_{\rm p}t\approx5300$ to $\omega_{\rm p}t\approx5800$, or from $\omega_{\rm p}t\approx6000$ to $\omega_{\rm p}t\approx8000$, as well as after the selected positron escapes upstream), the energy gain within the ``red'' half of the gyro-period is balanced by an approximately equal energy loss during the ``yellow'' half, since the background motional electric field $E_{\rm y,u}$ stays approximately constant throughout the  gyro-orbit. However, when the positron trajectory intersects the shock front, in the ``upstream'' portion of its clockwise Larmor gyration around the oblique field the positron $y$-velocity has always the same sign as the background motional electric field, thus resulting in systematic acceleration. Instead, the ``downstream'' portion of the orbit does not result in deceleration because in the simulation frame the motional electric field downstream is much smaller  than in the upstream.\footnote{We emphasize that a similar line of reasoning can be made in any other frame. In the particular frame of the simulation, the downstream motional electric field is $E_{\rm y,d}\ll E_{\rm y,u}$, which simplifies the argument. However, the shock is moving in this frame.} For electrons (not shown), whose Larmor gyration around the magnetic field is counter-clockwise, the $y$-velocity in the upstream medium will have opposite sign than $E_{\rm y,u}$, resulting again in net energization.

In summary, particles are systematically accelerated by the upstream motional electric field whenever their gyro-orbit intersects the shock surface, as expected for the Shock-Drift Acceleration process. SDA is inevitably associated with a $\grad B$ drift along (or opposite, according to the particle charge) the background electric field, due to the magnetic field jump   at the shock. This is seen, for the positron in Fig.~\fig{part15}, in the slow drift along $-\bf{\hat{y}}$ of $y(t)$ in panel \textit{\,c}. Such a drift will be more significant for higher obliquities, since the magnetic field jump at the shock is larger.

A comparison between $x(t)$ and $z(t)$ in Fig.~\fig{part15}\textit{\,c} suggests that our representative positron is heading upstream along the oblique magnetic field. As compared to a parallel shock, it will take longer to escape ahead of the shock, the more so for higher obliquities. Many gyro-orbits could then intersect the shock during a single shock-crossing of the positron gyro-center, thus resulting in multiple energizations (as  happens at $\ompt\approx5200$ and $\ompt\approx5900$). The particles that escape upstream are not systematically shock-drift accelerated any longer, until they return to the shock ($\omega_{\rm p}t\approx8000$ in Fig.~\fig{part15}), due to deflection by the upstream background fields or scattering off the upstream oblique waves. In that respect, even though the upstream motional field $E_{\rm y,u}$ is the primary agent for particle energization, DSA cooperates with SDA to trap the accelerated particles at the shock, thus increasing the overall acceleration efficiency. Making use of eq.~(\ref{eq:drift}) with a sample of $\sim1000$ high-energy particles, we are able to confirm that the ratio $\Delta\gamma_{\msc{sda}}/\Delta\gamma$ is independent of the total energy gain $\Delta\gamma$ \citep[as predicted by][]{jokipii_82}, and we estimate the fractional contribution by SDA to be $\sim60\%$. Therefore, for obliquity angles as small as $\theta=15\deg$, SDA is already the main acceleration mechanism.

\subsection{$\theta=30^\circ$}\label{sec:part30}
The acceleration process for a representative positron in a  shock with $\theta=30\deg$ is shown in Fig.~\fig{part30}. As the obliquity angle approaches the critical boundary between subluminal and superluminal configurations ($\theta_{\rm crit}\approx34\deg$), particles reflected upstream can hardly escape ahead of the shock. They are effectively trapped at the shock while heading upstream along the oblique magnetic field, since for $\theta\lesssim\thetacrit$ their gyro-center $x$-velocity is comparable to the shock speed. Before  escaping upstream, their orbit may intersect the shock many times (many more than for $\theta=15^\circ$), thus resulting in repeated energizations via SDA.

As discussed in \S\ref{sec:part15}, for a particle gyrating across the shock  any excursion into the upstream region results in efficient SDA by the upstream motional electric field $E_{\rm y,u}$ (compare panels \tit{a} and \tit{b} in Fig.~\fig{part30}), whereas in the downstream medium its energy stays roughly constant since $E_{\rm y,d}\ll E_{\rm y,u}$ in the simulation frame. The corresponding time evolution of the Lorentz factor $\gamma(t)$ resembles a stairway (see Fig.~\fig{part30}\textit{\,a}), with steps  that grow linearly with $\gamma$ since the time spent upstream is proportional to the particle energy. When a particle completes a full gyro-orbit in the upstream medium (between $\omega_{\rm p}t\approx7200$ and $\omega_{\rm p}t\approx7600$ for the positron in Fig.~\fig{part30}), its energy gain is balanced by an equal energy loss. Due to scattering off the incoming oblique waves or regular deflection by the upstream background fields, the particle may be re-captured by the shock, where its acceleration via SDA resumes. The representative positron in Fig.~\fig{part30}, as most of the high-energy particles extracted from our $\theta=30\deg$ simulation, is still being shock-accelerated at the final simulation time $\omega_{\rm p}t=9000$; its final Lorentz factor $\gamma\approx1000$, which is already much larger than in shocks with smaller obliquities, is therefore just a lower limit. 

Using eq.~(\ref{eq:drift}) with a sample of $\sim1000$ high-energy particles, we find that on average SDA contributes a fraction $\sim90\%$ to the overall energy gain, so that diffusive scattering plays a minor role in the acceleration process for $\theta=30\deg$. Thus, with increasing magnetic obliquity in the subluminal range, the acceleration process switches from DSA, which is the only mechanism acting for parallel shocks, to SDA, which for $\theta\lesssim\theta_{\rm crit}$ operates  in an essentially scatter-free regime. 

For $\theta=30\deg$ the upstream oblique waves are important to regulate injection into the acceleration process: all the particles that end up gaining the highest energies have been deflected from their upstream straight path before encountering the shock (see Fig.~\fig{part30}\textit{\,b} at $\ompt\approx6500$). We have tested that incoming particles enter the acceleration process only if their first shock encounter results in a reflection back upstream, whereas transmitted particles populate the low-energy bump of downstream spectra. In agreement with \citet{ballard_heavens_91}, we have verified that such reflected particles had encountered the shock with pitch angle cosine $\mu_{\msc{ht}}$ in the de Hoffmann-Teller frame \citep{deHoffmann_Teller_50} which satisfies 
\begin{equation}\label{eq:inj}
|\mu_{\msc{ht}}|<\sqrt{1-\frac{B_{\rm u,\msc{ht}}}{B_{\rm d,\msc{ht}}}}
\end{equation}
 where $B_{\rm d,\msc{ht}}/B_{\rm u,\msc{ht}}$ is the magnetic field jump in the de Hoffmann-Teller frame. In other words, only the incoming particles that are deflected by the upstream waves, such that their pitch angle satisfies \eq{inj} when encountering the shock, will be reflected back upstream and participate in the SDA process.

In summary, while for quasi-parallel shocks ($\theta\lesssim10\deg$) the upstream oblique waves generated by the beam of returning particles provide the upstream scattering for DSA, for magnetic obliquities close to $\theta_{\rm crit}$ the same waves mediate the injection process into SDA.

\section{Discussion and Summary}\label{sec:disc}
We have explored by means of 2.5D PIC simulations the internal structure and acceleration properties of relativistic collisionless shocks propagating in a magnetized pair plasma. A strong magnetic field may significantly suppress cross-field motion and therefore quench the growth of the Weibel instability \citep{weibel_59, medvedev_loeb_99,gruzinov_waxman_99}, which is believed to mediate unmagnetized or weakly magnetized shocks \citep[e.g.,][]{spitkovsky_05, spitkovsky_08, spitkovsky_08b,chang_08,keshet_08}. Instead, strongly magnetized perpendicular shocks are triggered by magnetic reflection of the incoming particles from the shock-compressed fields \citep[e.g.,][]{alsop_arons_88}. \citet{spitkovsky_05} finds that the transition between weakly magnetized (Weibel-mediated) and strongly magnetized (reflection-mediated) perpendicular shocks happens around an upstream magnetization $\sigma\sim10^{-3}$. In this work, we focus on strongly magnetized upstream flows ($\sigma=0.1$) and we study the dependence of the shock properties on the inclination angle $\theta$ between the upstream magnetic field and the shock normal, as measured in the ``wall'' frame of our simulations.

In magnetized shocks,  particle gyro-centers are constrained to move along the magnetic field. It is then natural to define a critical obliquity angle $\thetacrit$ such that in ``superluminal'' shocks   ($\theta>\thetacrit$) particles cannot escape upstream along the magnetic field lines. If the upstream plasma has Lorentz factor $\gamma_0=15$ and magnetization $\sigma=0.1$, MHD calculations yield $\theta_{\rm crit}\approx34\deg$ in the simulation frame, weakly dependent on both $\gamma_0$ and $\sigma$.  Since for $\theta>\thetacrit$ particles are quickly advected downstream from the shock, the acceleration efficiency should be strongly suppressed for superluminal configurations. In fact, we find that magnetized relativistic pair shocks produce a significant population of suprathermal particles only for ``subluminal'' magnetic geometries ($\theta<\thetacrit$). This holds true for a wide range of upstream Lorentz factors (we also tried $\gamma_0=5$ and $\gamma_0=50$, for fixed $\sigma=0.1$) and magnetizations (we also tested $\sigma=0.03$ and $\sigma=0.3$, for fixed $\gamma_0=15$). We conclude that only a narrow range of obliquity angles allows for efficient particle acceleration in relativistic ($\gamma_0\gtrsim5$) magnetized ($\sigma\gtrsim0.03$) pair shocks: as measured from the upstream frame, the magnetic field should lie within a cone of half-opening angle $\approx34\deg/\gamma_0$ around the shock normal to produce a significant population of nonthermal particles.\footnote{We remark that this conclusion applies only to electron-positron shocks. For electron-ion magnetized shocks, the electric potential across the shock complicates the particle dynamics and we cannot exclude that shocks which are nominally superluminal may be moderate particle accelerators. Relativistic shocks propagating in magnetized electron-ion plasmas will be discussed in a forthcoming paper.}

For quasi-parallel magnetic configurations ($\theta\lesssim10\deg$), some of the incoming particles are scattered back upstream and then gain energy by bouncing between the shock and the upstream medium. The upstream scattering is provided by oblique waves which propagate towards the shock and are self-consistently triggered by the  shock-accelerated particles that escape upstream; magnetic turbulence in the shock transition layer reflects the high-energy particles back upstream. The stochastic nature of the acceleration process suggests that DSA is the main energization mechanism in quasi-parallel magnetized pair shocks, at the very least because Shock-Surfing Acceleration does not operate in electron-positron shocks and the absence of a background motional electric field rules out SDA in parallel shocks. For $\theta=0\deg$, we find that at the final time  of our simulation ($\ompt=9000$) the suprathermal tail of the downstream energy spectrum accounts for $\sim1\%$ of the downstream particle number and $\sim4\%$ of the energy, with a power-law spectral index $-2.8\pm0.1$; however, these may be lower limits, since further evolution of the spectrum in our simulations of quasi-parallel shocks may be artificially suppressed by our choice of boundary conditions.

Whereas for quasi-parallel magnetic geometries ($\theta\lesssim10\deg$) the shock-generated turbulent fields are typically stronger than the uniform motional electric field, and so SDA plays a minor role in the acceleration process, its mean fractional contribution to the particle overall energy gain increases to $\sim60\%$ for $\theta=15\deg$ and reaches $\sim90\%$ for $\theta=30\deg$. Therefore, the acceleration process for $\theta\lesssim\thetacrit$ is almost scatter-free: particles heading upstream along the oblique magnetic field hardly escape from the shock, and they can experience multiple shock-drift energizations as their orbit repeatedly crosses the shock surface. However, we cannot exclude that a fraction of the shock-drift accelerated particles may eventually gain enough energy to become effectively ``unmagnetized'', i.e., their gyrocenter would no longer be constrained to follow the background field. Such particles may be further accelerated by a diffusive Fermi-like mechanism, and the SDA process would then serve as a pre-acceleration stage before the onset of DSA. The upstream oblique waves, which provide the upstream scattering in quasi-parallel shocks, may also mediate this further step of diffusive acceleration for larger, but still subluminal, obliquities. We also find that in $\theta\lesssim\thetacrit$ shocks  these waves regulate the process of particle injection into SDA. 

As predicted by \citet{jokipii_82}, we find that SDA in $\theta\lesssim\theta_{\rm crit}$ shocks is a much faster process than DSA in quasi-parallel shocks, since each SDA cycle takes approximately one gyro-period, which is much shorter than any diffusive scattering timescale. As the acceleration mechanism switches from DSA to SDA with increasing obliquity, the  acceleration efficiency increases as well. For $\theta=30\deg$, the number and energy fractions contributed by the suprathermal tail of the downstream particle spectrum are respectively $\sim2\%$ and $\sim12\%$ and its power-law spectral index is $-2.3\pm0.1$. Since acceleration in $\theta\approx\thetacrit$ shocks is close to a steady state around the final time of our simulations ($\ompt=9000$), the efficiencies quoted above should be near the saturation values.

For superluminal shocks ($\theta\gtrsim\theta_{\rm crit}$), we find that if the upstream background field is not highly turbulent on scales smaller than the transverse size  of our simulations ($\sim100\comp$), the self-generated turbulence is not strong enough to trap a significant fraction of particles at the shock for efficient acceleration; rather, most of the particles are quickly advected downstream with the magnetic field lines. 
Monte-Carlo simulations of relativistic magnetized shocks found evidence for particle acceleration in perpendicular shocks \citep[e.g.,][]{ostrowski_bednarz_98,niemiec_ostrowski_04,ellison_double_04}, but they required exceptionally high levels of turbulence \citep{ostrowski_bednarz_02}. We show via PIC simulations with  laminar upstream fields that superluminal shocks do not self-consistently generate such  large degrees of turbulence, and they are inefficient particle accelerators. In Appendix B we show that the drop in acceleration efficiency at $\thetacrit\approx34\deg$ between subluminal and superluminal shocks is rather abrupt.

We have tested that our results do not significantly depend on the number of particles per computational  cell (we tried 2 and 8 particles per species per cell) or the transverse size of our two-dimensional simulation box (we tried 512, 1024, 2048 and 4096 transverse cells). We have verified that the geometry of the magnetic field with respect to the simulation plane (in-plane or out-of-plane magnetic field) does not change our main conclusions (see Appendix A); moreover, our 2.5D results have been confirmed by running a limited suite of simulations in three-dimensional domains with 64, 128 or 256 cells along each transverse dimension. 

A direct comparison of 2.5D simulations with 3D runs is possible, since for moderate obliquities ($\theta\lesssim45\deg$) the gyro-orbits of downstream particles mostly lie outside the two-dimensional simulation plane, and the flow thermalizes behind the shock to a \tit{three-dimensional}  distribution. For the obliquity angles investigated with 3D simulations ($\theta=30\deg$ and $\theta=35\deg$), we see a remarkable agreement with the results of our 2.5D runs (for $\theta=30\deg$, see Appendix A). \citet{jones_98} found that numerical simulations in two spatial dimensions could artificially inhibit particle diffusion across field lines, which is required for efficient particle acceleration in superluminal shocks. Instead, we confirm with 3D simulations that  superluminal relativistic shocks propagating in pair plasmas with laminar fields do not produce a significant population of nonthermal particles, at least for the domain sizes we tried in 3D. 

In summary, our results underscore that particle acceleration in relativistic ($\gamma_0\gtrsim5$) magnetized ($\sigma\gtrsim0.03$) pair flows is efficient only for magnetic obliquities $\theta\lesssim\theta_{\rm crit}\approx34\deg$. As seen from the upstream fluid frame, only nearly-parallel shocks (with obliquity $\lesssim34\deg/\gamma_0$) produce a significant suprathermal tail of accelerated particles. These findings place constraints on the models of Pulsar Wind Nebulae (PWNe), Active Galactic Nuclei (AGN) jets and Gamma-Ray Bursts (GRBs) that invoke particle acceleration in relativistic magnetized  shocks.

Recent relativistic MHD simulations of PWNe \citep{komissarov_03, komissarov_04,delzanna_04} have shown that polar jets with the observed velocities  can develop provided that the magnetization of the relativistic ($\gamma_0\gtrsim10^6$) electron-positron pulsar wind is large enough ($\sigma\gtrsim0.01$) to collimate the jet by hoop stresses downstream of the wind termination shock. Since the asymptotic wind magnetic field is mostly toroidal around the pulsar axis, the termination shock will be quasi-perpendicular. If the wind magnetization is $\sigma\gtrsim0.03$, our results predict that the termination shock should not produce a significant population of nonthermal  particles, which is required instead by the observed synchrotron and inverse Compton emission. However, for oblique rotators, the pulsar wind in the equatorial plane should consist of stripes of alternating field, and the  field may dissipate via magnetic reconnection of the opposite polarity stripes before the termination shock \citep{coroniti_90,lyubarsky_kirk_01}; the effective magnetization of the shock would then be lower and particle acceleration would proceed like in an essentially unmagnetized shock. Alternatively, magnetic reconnection of the striped wind field when compressed at the termination shock may be responsible for particle energization \citep{petri_lyubarsky_07,lyubarsky_liverts_08}. 

Synchrotron and inverse Compton emission from the core of blazar jets is usually attributed to high-energy electrons and positrons accelerated in mildly-relativistic internal shocks ($\gamma_0\sim2$ in the jet comoving frame). Such shocks should be quasi-perpendicular, since polarization measurements of radio knots \citep{gabuzda_04, pushkarev_05} indicate that the magnetic field is mostly transverse to the jet axis. If the jet flow is significantly magnetized ($\sigma\gtrsim0.03$), the presence of a substantial population of high-energy particles accelerated at such perpendicular magnetized shocks would be hard to explain, unless particle acceleration proceeds not in the central ``spine'' of the jet but rather at its edges; here, interaction of the jet with the surrounding medium may create subluminal configurations where particle acceleration would be more efficient. Otherwise, the jet flow should be weakly magnetized ($\sigma\lesssim0.03$). Alternatively, particle acceleration may occur not in shocks but in reconnection layers of a Poynting-dominated jet, although \citet{sikora_05} have shown that the jet is likely to be matter-dominated at the pc-scale emission region.

Similarly, in the standard ``fireball'' scenario of GRBs, most models of the prompt emission require a population of energetic particles accelerated at mildly-relativistic internal shocks. If the ejecta are strongly magnetized, possibly by the field of the progenitor star, and the magnetic field is mostly transverse to the outflow, we would predict inefficient particle acceleration and thus negligible nonthermal emission, contrary to observations. This could be circumvented if the relativistic outflow primarily takes the form of a large-scale Poynting flux, and particle acceleration occurs not in shocks but by dissipation of magnetic energy via current-driven instabilities \citep{lyutikov_03}.

In summary, our results suggest that models of nonthermal emission in PWNe, AGN jets and GRBs which require particle acceleration in relativistic pair shocks should resort to either nearly-parallel or weakly magnetized shocks. Alternatively, other acceleration mechanisms, e.g., by dissipation of the field via magnetic reconnection, should be invoked, or the upstream flow should contain a significant ionic component \citep{hoshino_92,amato_arons_06} or small-scale turbulence \citep{sironi_goodman_07}. 

\acknowledgements{We thank J. Arons, M. Baring, M. Dieckmann, D. Eichler, D. Ellison, J. Goodman, U. Keshet and H. Spruit for useful comments and suggestions. We thank S. Komissarov for providing us with the solutions of relativistic MHD Riemann problem for comparison with PIC results. This research was supported by NSF grant AST-0807381 and BSF grant
2006095. A.S. acknowledges the support from Alfred P. Sloan Foundation
fellowship.}

\bibliography{accel}

\begin{thebibliography}{53}
\expandafter\ifx\csname natexlab\endcsname\relax\def\natexlab#1{#1}\fi

\bibitem[{{Achterberg} {et~al.}(2001){Achterberg}, {Gallant}, {Kirk}, \&
  {Guthmann}}]{achterberg_01}
{Achterberg}, A., {Gallant}, Y.~A., {Kirk}, J.~G., \& {Guthmann}, A.~W. 2001,
  \mnras, 328, 393

\bibitem[{{Alsop} \& {Arons}(1988)}]{alsop_arons_88}
{Alsop}, D. \& {Arons}, J. 1988, Physics of Fluids, 31, 839

\bibitem[{{Amato} \& {Arons}(2006)}]{amato_arons_06}
{Amato}, E. \& {Arons}, J. 2006, \apj, 653, 325

\bibitem[{{Appl} \& {Camenzind}(1988)}]{appl_camenzind_88}
{Appl}, S. \& {Camenzind}, M. 1988, \aap, 206, 258

\bibitem[{{Ballard} \& {Heavens}(1991)}]{ballard_heavens_91}
{Ballard}, K.~R. \& {Heavens}, A.~F. 1991, \mnras, 251, 438

\bibitem[{{Bednarz} \& {Ostrowski}(1998)}]{ostrowski_bednarz_98}
{Bednarz}, J. \& {Ostrowski}, M. 1998, Physical Review Letters, 80, 3911

\bibitem[{{Begelman} \& {Kirk}(1990)}]{begelman_kirk_90}
{Begelman}, M.~C. \& {Kirk}, J.~G. 1990, \apj, 353, 66

\bibitem[{{Bell}(1978)}]{bell_78}
{Bell}, A.~R. 1978, \mnras, 182, 147

\bibitem[{{Blandford} \& {Eichler}(1987)}]{blandford_eichler_87}
{Blandford}, R. \& {Eichler}, D. 1987, \physrep, 154, 1

\bibitem[{{Blandford} \& {Ostriker}(1978)}]{blandford_ostriker_78}
{Blandford}, R.~D. \& {Ostriker}, J.~P. 1978, \apjl, 221, L29

\bibitem[{{Bret} \& {Dieckmann}(2008)}]{bret_08}
{Bret}, A. \& {Dieckmann}, M.~E. 2008, Physics of Plasmas, 15, 062102

\bibitem[{{Buneman}(1993)}]{buneman_93}
{Buneman}, O. 1993, {in ``Computer Space Plasma Physics'', Terra Scientific,
  Tokyo, 67}

\bibitem[{{Chang} {et~al.}(2008){Chang}, {Spitkovsky}, \& {Arons}}]{chang_08}
{Chang}, P., {Spitkovsky}, A., \& {Arons}, J. 2008, \apj, 674, 378

\bibitem[{{Chen} \& {Armstrong}(1975)}]{chen_75}
{Chen}, G. \& {Armstrong}, T.~P. 1975, in International Cosmic Ray Conference,
  Vol.~5, 1814--1819

\bibitem[{{Coroniti}(1990)}]{coroniti_90}
{Coroniti}, F.~V. 1990, \apj, 349, 538

\bibitem[{{de Hoffmann} \& {Teller}(1950)}]{deHoffmann_Teller_50}
{de Hoffmann}, F. \& {Teller}, E. 1950, Physical Review, 80, 692

\bibitem[{{Del Zanna} {et~al.}(2004){Del Zanna}, {Amato}, \&
  {Bucciantini}}]{delzanna_04}
{Del Zanna}, L., {Amato}, E., \& {Bucciantini}, N. 2004, \aap, 421, 1063

\bibitem[{{Drury}(1983)}]{drury_83}
{Drury}, L.~O. 1983, Reports on Progress in Physics, 46, 973

\bibitem[{{Ellison} \& {Double}(2004)}]{ellison_double_04}
{Ellison}, D.~C. \& {Double}, G.~P. 2004, Astroparticle Physics, 22, 323

\bibitem[{{Gabuzda} {et~al.}(2004){Gabuzda}, {Murray}, \&
  {Cronin}}]{gabuzda_04}
{Gabuzda}, D.~C., {Murray}, {\'E}., \& {Cronin}, P. 2004, \mnras, 351, L89

\bibitem[{{Gallant} {et~al.}(1992){Gallant}, {Hoshino}, {Langdon}, {Arons}, \&
  {Max}}]{gallant_92}
{Gallant}, Y.~A., {Hoshino}, M., {Langdon}, A.~B., {Arons}, J., \& {Max}, C.~E.
  1992, \apj, 391, 73

\bibitem[{Greenwood {et~al.}(2004)Greenwood, Cartwright, Luginsland, \&
  Baca}]{greenwood_04}
Greenwood, A.~D., Cartwright, K.~L., Luginsland, J.~W., \& Baca, E.~A. 2004,
  Journal of Computational Physics, 201, 665

\bibitem[{{Gruzinov} \& {Waxman}(1999)}]{gruzinov_waxman_99}
{Gruzinov}, A. \& {Waxman}, E. 1999, \apj, 511, 852

\bibitem[{{Hoshino}(2008)}]{hoshino_08}
{Hoshino}, M. 2008, \apj, 672, 940

\bibitem[{{Hoshino} \& {Arons}(1991)}]{hoshino_91}
{Hoshino}, M. \& {Arons}, J. 1991, Physics of Fluids B, 3, 818

\bibitem[{{Hoshino} {et~al.}(1992){Hoshino}, {Arons}, {Gallant}, \&
  {Langdon}}]{hoshino_92}
{Hoshino}, M., {Arons}, J., {Gallant}, Y.~A., \& {Langdon}, A.~B. 1992, \apj,
  390, 454

\bibitem[{{Jokipii}(1982)}]{jokipii_82}
{Jokipii}, J.~R. 1982, \apj, 255, 716

\bibitem[{{Jones} {et~al.}(1998){Jones}, {Jokipii}, \& {Baring}}]{jones_98}
{Jones}, F.~C., {Jokipii}, J.~R., \& {Baring}, M.~G. 1998, \apj, 509, 238

\bibitem[{{Keshet} {et~al.}(2008){Keshet}, {Katz}, {Spitkovsky}, \&
  {Waxman}}]{keshet_08}
{Keshet}, U., {Katz}, B., {Spitkovsky}, A., \& {Waxman}, E. 2008,
  ArXiv:astro-ph/0802.3217

\bibitem[{{Keshet} \& {Waxman}(2005)}]{keshet_waxman_05}
{Keshet}, U. \& {Waxman}, E. 2005, Physical Review Letters, 94, 111102

\bibitem[{{Kirk} {et~al.}(2000){Kirk}, {Guthmann}, {Gallant}, \&
  {Achterberg}}]{kirk_00}
{Kirk}, J.~G., {Guthmann}, A.~W., {Gallant}, Y.~A., \& {Achterberg}, A. 2000,
  \apj, 542, 235

\bibitem[{{Kirk} \& {Heavens}(1989)}]{kirk_heavens_89}
{Kirk}, J.~G. \& {Heavens}, A.~F. 1989, \mnras, 239, 995

\bibitem[{{Komissarov} \& {Lyubarsky}(2003)}]{komissarov_03}
{Komissarov}, S.~S. \& {Lyubarsky}, Y.~E. 2003, \mnras, 344, L93

\bibitem[{{Komissarov} \& {Lyubarsky}(2004)}]{komissarov_04}
---. 2004, \mnras, 349, 779

\bibitem[{{Landau} {et~al.}(1984){Landau}, {Pitaevskii}, \&
  {Lifshitz}}]{landau_60}
{Landau}, L.~D., {Pitaevskii}, L.~P., \& {Lifshitz}, E.~M. 1984,
  {Electrodynamics of continuous media} (Elsevier Science \& Technology Books)

\bibitem[{{Langdon} {et~al.}(1988){Langdon}, {Arons}, \& {Max}}]{langdon_88}
{Langdon}, A.~B., {Arons}, J., \& {Max}, C.~E. 1988, Physical Review Letters,
  61, 779

\bibitem[{{Lee} {et~al.}(1996){Lee}, {Shapiro}, \& {Sagdeev}}]{lee_96}
{Lee}, M.~A., {Shapiro}, V.~D., \& {Sagdeev}, R.~Z. 1996, \jgr, 101, 4777

\bibitem[{{Lyubarsky} \& {Kirk}(2001)}]{lyubarsky_kirk_01}
{Lyubarsky}, Y. \& {Kirk}, J.~G. 2001, \apj, 547, 437

\bibitem[{{Lyubarsky} \& {Liverts}(2008)}]{lyubarsky_liverts_08}
{Lyubarsky}, Y. \& {Liverts}, M. 2008, \apj, 682, 1436

\bibitem[{{Lyutikov} \& {Blandford}(2003)}]{lyutikov_03}
{Lyutikov}, M. \& {Blandford}, R. 2003, ArXiv:astro-ph/0312347

\bibitem[{{Medvedev} \& {Loeb}(1999)}]{medvedev_loeb_99}
{Medvedev}, M.~V. \& {Loeb}, A. 1999, \apj, 526, 697

\bibitem[{{Niemiec} \& {Ostrowski}(2004)}]{niemiec_ostrowski_04}
{Niemiec}, J. \& {Ostrowski}, M. 2004, \apj, 610, 851

\bibitem[{{Ostrowski} \& {Bednarz}(2002)}]{ostrowski_bednarz_02}
{Ostrowski}, M. \& {Bednarz}, J. 2002, \aap, 394, 1141

\bibitem[{{P{\'e}tri} \& {Lyubarsky}(2007)}]{petri_lyubarsky_07}
{P{\'e}tri}, J. \& {Lyubarsky}, Y. 2007, \aap, 473, 683

\bibitem[{{Pushkarev} {et~al.}(2005){Pushkarev}, {Gabuzda}, {Vetukhnovskaya},
  \& {Yakimov}}]{pushkarev_05}
{Pushkarev}, A.~B., {Gabuzda}, D.~C., {Vetukhnovskaya}, Y.~N., \& {Yakimov},
  V.~E. 2005, \mnras, 356, 859

\bibitem[{{Sikora} {et~al.}(2005){Sikora}, {Begelman}, {Madejski}, \&
  {Lasota}}]{sikora_05}
{Sikora}, M., {Begelman}, M.~C., {Madejski}, G.~M., \& {Lasota}, J.-P. 2005,
  \apj, 625, 72

\bibitem[{{Sironi} \& {Goodman}(2007)}]{sironi_goodman_07}
{Sironi}, L. \& {Goodman}, J. 2007, \apj, 671, 1858

\bibitem[{{Spitkovsky}(2005)}]{spitkovsky_05}
{Spitkovsky}, A. 2005, {in AIP Conference Series, Vol.~801, 345-350,
  astro-ph/0603211}

\bibitem[{{Spitkovsky}(2008{\natexlab{a}})}]{spitkovsky_08}
---. 2008{\natexlab{a}}, \apjl, 673, L39

\bibitem[{{Spitkovsky}(2008{\natexlab{b}})}]{spitkovsky_08b}
---. 2008{\natexlab{b}}, \apjl, 682, L5

\bibitem[{{Synge}(1957)}]{synge_57}
{Synge}, J.~L. 1957, {The Relativistic Gas (North-Holland, Amsterdam), 33}

\bibitem[{{Webb} {et~al.}(1983){Webb}, {Axford}, \& {Terasawa}}]{webb_83}
{Webb}, G.~M., {Axford}, W.~I., \& {Terasawa}, T. 1983, \apj, 270, 537

\bibitem[{{Weibel}(1959)}]{weibel_59}
{Weibel}, E.~S. 1959, Physical Review Letters, 2, 83

\end{thebibliography}

\appendix
\section{A) 2.5D simulations with in-plane upstream background magnetic field}\label{append1}
In the 2.5D simulations presented in this work, the fact that the upstream background magnetic field lies in a plane which is perpendicular to the simulation plane (``out-of-plane'' magnetic configurations) may be suspected in reducing turbulence along the field. Here we test this by analyzing the results of 2.5D simulations with an ``in-plane'' upstream blackground magnetic field, for each of the subluminal oblique configurations presented in \S \ref{sec:struct}, i.e., $\theta=15\deg$ and $\theta=30\deg$.

In the in-plane case, the upstream background magnetic field will have components $B_{\rm x,u}\geq0$ and $B_{\rm y,u}\geq0$ (so that $\theta\equiv\arctan [B_{\rm y,u}/B_{\rm x,u}]$), and the associated upstream motional electric field will be $\mathbf{E}_{\rm u}=\beta_0B_{\rm y,u}\,\hat{\mathbf{z}}$. As regards to the electromagnetic fields, the direction $+\hat{\mathbf{z}}$ in out-of-plane configurations should then correspond to in-plane $+\hat{\mathbf{y}}$, and out-of-plane $+\hat{\mathbf{y}}$ to in-plane $-\hat{\mathbf{z}}$. Apart from this overall rotation of the coordinate system, we expect out-of-plane and in-plane geometries with the same magnetic obliquity to show comparable results in terms of  shock structure and particle spectra. Any difference should be an artifact of the reduced dimensionality of our simulation box and would be a caveat against the generalization of our 2.5D results to three-dimensional problems. 

\begin{figure*}[bp]
\begin{center}
\includegraphics[width=0.6\textwidth]{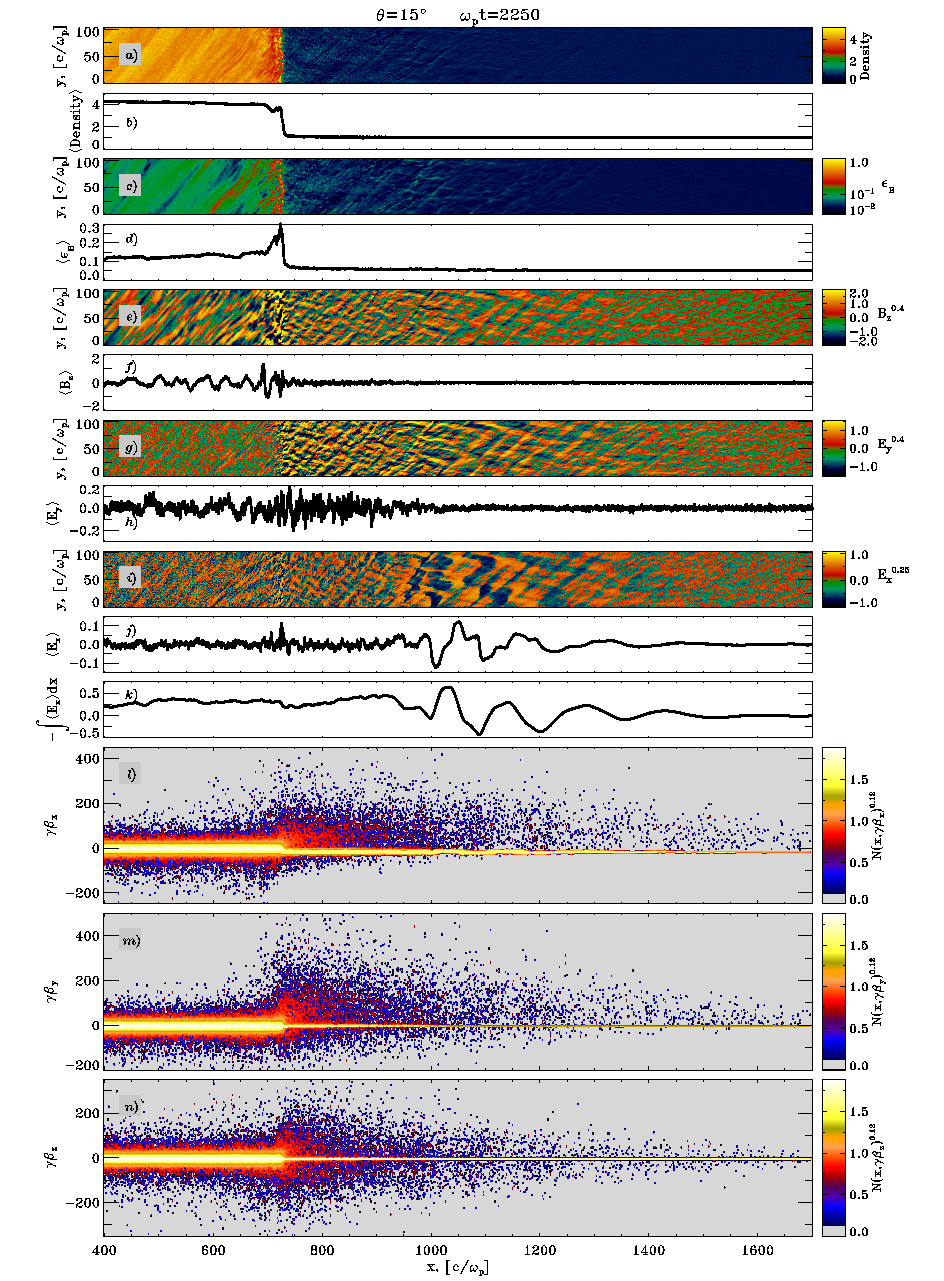}
\caption{Shock structure and positron phase space at time $\omega_{\rm{p}}t=2250$ for $\theta=15^\circ$, with upstream background magnetic field lying in the simulation plane. The fluid quantities are normalized as in Fig.~\fig{fluid0}, but here $B_{\rm z}$ and $E_{\rm y}$ are in units of the transverse component $B_{\rm{y,u}}$ of the upstream background magnetic field.}
\label{fig:fluid15-90}
\end{center}
\end{figure*}

\begin{figure*}[htbp]
\begin{center}
\includegraphics[width=0.6\textwidth]{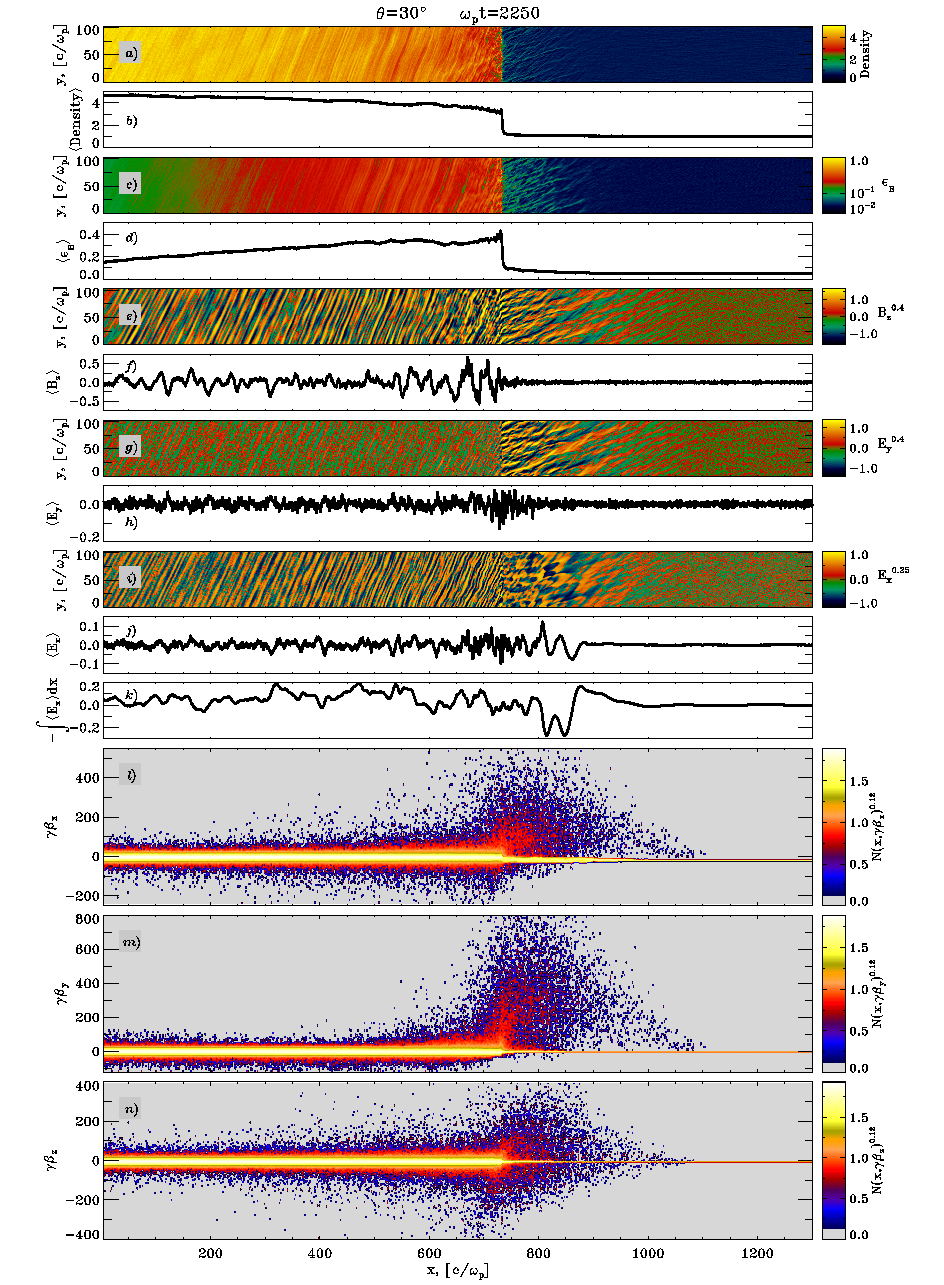}
\caption{Shock structure and positron phase space at time $\omega_{\rm{p}}t=2250$ for $\theta=30^\circ$, with upstream background magnetic field lying in the simulation plane. The fluid quantities are normalized as in Fig.~\fig{fluid15-90}.}
\label{fig:fluid30-90}
\end{center}
\end{figure*}

\subsection{Shock structure}
Figs.~\fig{fluid15-90} and \fig{fluid30-90} show the shock structure and positron phase space at time $\ompt=2250$ for in-plane magnetic configurations with obliquity $\theta=15\deg$ and $\theta=30\deg$ respectively. Their out-of-plane counterparts are plotted  in Figs.~\fig{fluid15} and \fig{fluid30} respectively. We find that the shock internal structure does not significantly change by varying the orientation of the field with respect to the simulation plane. 

Out-of-plane and in-plane configurations with the same obliquity show a comparable shock  velocity and similar transversely-averaged density (panel \textit{b}) and magnetic energy (panel  \textit{d}) profiles, as relates to the width of the shock transition region ($\sim10\comp$), the jump across the shock (in agreement with three-dimensional MHD calculations) and the presence of a fast and a slow shock (compare Fig.~\fig{fluid30} and Fig.~\fig{fluid30-90} for $\theta=30\deg$). For $\theta=15\deg$, the $y$-averaged magnetic energy profile peaks in the shock layer at $\sim30\%$ of the upstream kinetic energy density, both for out-of-plane (Fig.~\fig{fluid15}\pan{d}) and in-plane (Fig.~\fig{fluid15-90}\pan{d}) configurations, and in both cases it decays downstream  on a lenghtscale $\sim50\comp$. For $\theta=30\deg$, the overall fluid structure does not significantly change between in-plane and out-of-plane configurations, but a few minor differences are present: for the in-plane case, the magnetic energy in the shock layer slightly exceeds the downstream plateau  (Fig.~\fig{fluid30-90}\textit{\,d}), which was not observed in its out-of-plane counterpart (Fig.~\fig{fluid30}\textit{\,d}); also, the speed of the slow shock, tracked from the point where the magnetic energy starts to decrease downstream from the  main shock, seems to be higher for in-plane than  out-of-plane configurations. For both $\theta=15\deg$ and $\theta=30\deg$, the motional electric field (panel \tit{h} in Figs.~\fig{fluid15} and \fig{fluid30}, not shown in Figs.~\fig{fluid15-90} and \fig{fluid30-90}) does not drop to zero behind the main shock due to a residual downstream bulk velocity transverse to the upstream flow ($\beta_{\rm y,d}<0$ and $\beta_{\rm z,d}<0$ for in-plane and out-of-plane configurations respectively). The transient electromagnetic precursor wave generated by coherent Larmor bunching at the shock, which showed up in  $B_{\rm z}$ and $E_{\rm y}$ for out-of-plane oblique configurations, is now seen in the corresponding in-plane fields $B_{\rm y}$ and $E_{\rm z}$ with comparable wavelengths and amplitude (further upstream than Figs.~\fig{fluid15-90} and \fig{fluid30-90} show). 

The phase space plots  in panels \textit{l-n} of Figs.~\fig{fluid15-90} and \fig{fluid30-90} are similar to their counterparts in Figs.~\fig{fluid15} and \fig{fluid30} respectively, provided that out-of-plane $\gamma\beta_{\rm z}$ and $\gamma\beta_{\rm y}$ are compared with in-plane $\gamma\beta_{\rm y}$ and $-\gamma\beta_{\rm z}$ respectively. As observed for out-of-plane configurations, the returning high-energy particles slide along the upstream magnetic field, which means for the in-plane case that they show large and positive $\gamma\beta_{\rm x}$ and $\gamma\beta_{\rm y}$ (panels \tit{l} and \tit{m} in Figs.~\fig{fluid15-90},\,\fig{fluid30-90}). The in-plane excess of positrons with $\gamma\beta_{\rm z}>0$ in the upstream region close to the shock (panel \textit{n} in Figs.~\fig{fluid15-90} and \fig{fluid30-90}) is due to $\grad B$ drift, and it corresponds to the asymmetry towards negative values of $\gamma\beta_{\rm y}$ (panel \tit{m} in Figs.~\fig{fluid15} and \fig{fluid30} respectively) observed for out-of-plane configurations. 

Aside from the similarities between in-plane and out-of-plane results, Figs.~\fig{fluid15-90} and \fig{fluid30-90} improve our knowledge of the internal structure of oblique shocks. The two-dimensional plots of magnetic energy density (panel \textit{c}) and $B_{\rm z}$ (panel \tit{e}) of Figs.~\fig{fluid15} and \fig{fluid30} show islands of low magnetic field in the downstream region of the strong shock. Fewer islands are produced at the shock at later times; when formed, each island slowly decays with time at constant $x$-location, and it disappears when caught up by the slow shock. The corresponding panels in Figs.~\fig{fluid15-90} and \fig{fluid30-90} clarify that such islands are actually tubes of low magnetic energy stretching along the downstream magnetic field, as confirmed also by 3D simulations with 256  cells along each transverse dimension.

Most importantly, the electromagnetic component of the upstream oblique waves, which was mostly seen in $B_{\rm z}$ and $E_{\rm y}$ for the out-of-plane case (panels \textit{e,\,g} in Figs.~\fig{fluid15},\,\fig{fluid30}), shows up mainly in the same field components also for in-plane configurations. In other words, the wave magnetic field is still mostly perpendicular to the simulation plane, regardless of the orientation of the upstream background magnetic field  with respect to the simulation plane. A further study of such modes will therefore have to be fully three-dimensional to capture their true physical nature. What Figs.~\fig{fluid15-90} and \fig{fluid30-90} (panels \textit{\,e,\,g,\,i}) add to our understanding of such waves is that, in the region far upstream where they are triggered, the surface of constant phase is roughly a cone of constant opening angle around the upstream background field.

\subsection{Particle energy spectra}
\begin{figure}
\centering
\subfigure[For fixed magnetic obliquity $\theta=15^\circ$, downstream particle spectra at time $\omega_{\rm{p}}t=9000$ for different magnetic field geometries: magnetic field lying either in the simulation plane (black) or in a plane perpendicular to the simulation plane (red).]{
\includegraphics[width=0.47\textwidth]{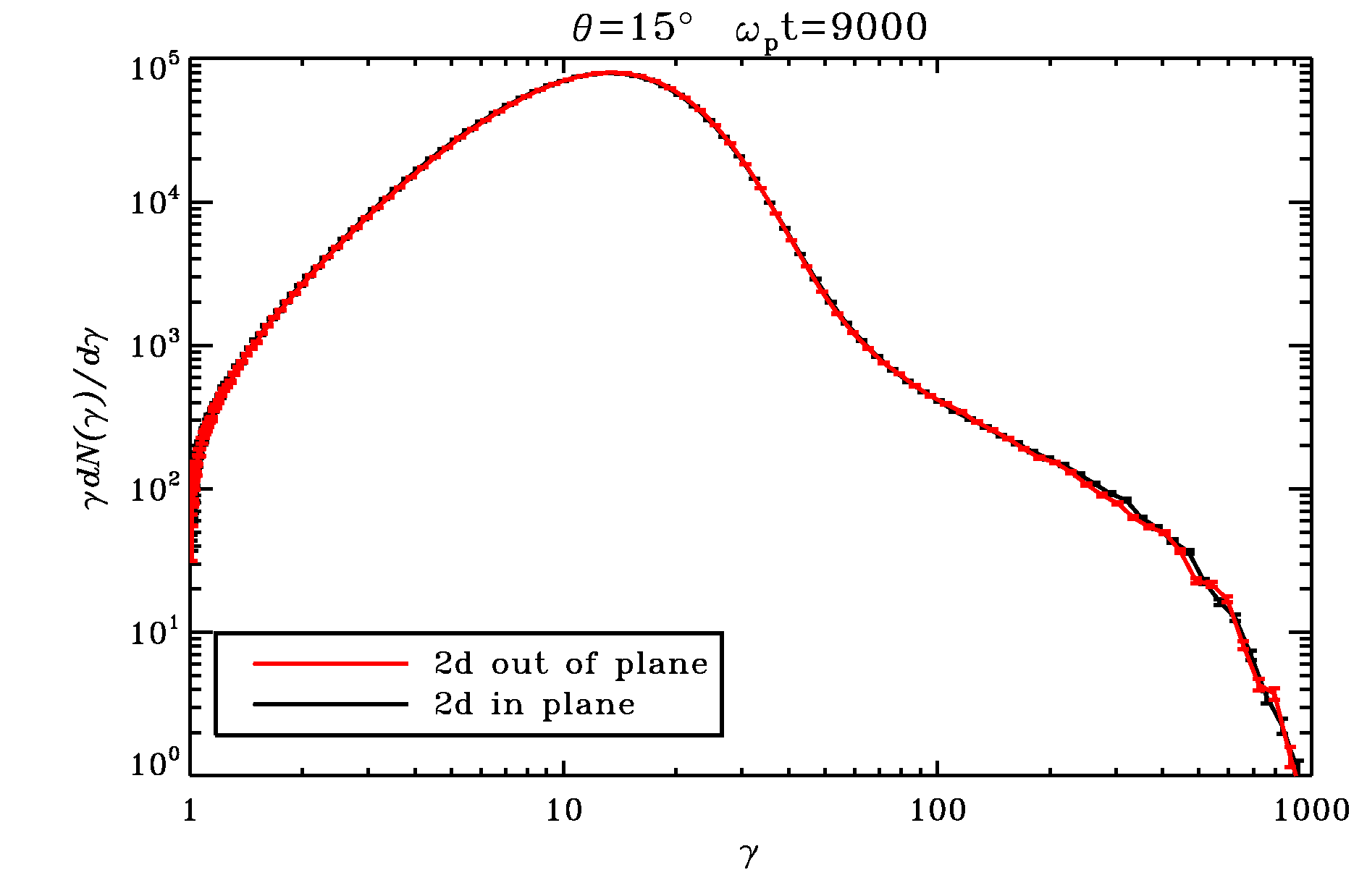}}
\hspace{0.1in}
\subfigure[For fixed magnetic obliquity $\theta=30^\circ$, downstream particle spectra at time $\omega_{\rm{p}}t=9000$ for different magnetic field geometries: two-dimensional simulations with either in-plane (black) or out-of-plane (red) magnetic field and a three-dimensional run (green).]{
\includegraphics[width=0.47\textwidth]{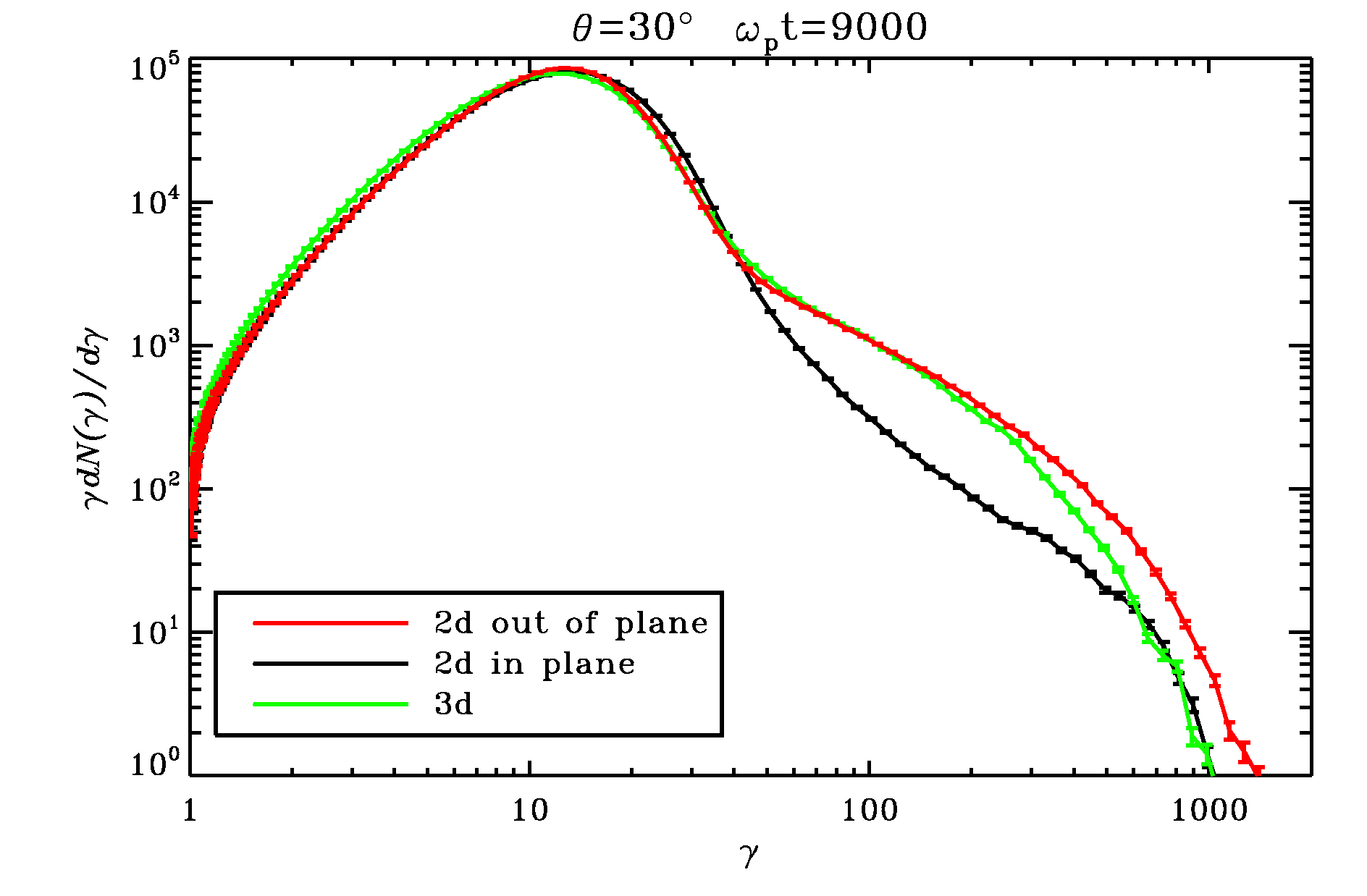}}
\caption{Comparison of downstream particle spectra among different geometries of the upstream background magnetic field, for magnetic obliquity $\theta=15\deg$ in panel (a) and $\theta=30\deg$ in panel (b).}
\label{fig:specphi}
\end{figure}

Fig.~\fig{specphi} shows at $\ompt=9000$ a comparison of downstream particle spectra between in-plane (black line) and out-of-plane (red line) configurations: for $\theta=15\deg$ (panel \tit{a}), the spectrum is remarkably similar between the two cases, whereas a significant difference is seen for  $\theta=30\deg$ (panel \tit{b}), with the high-energy tail for the out-of-plane configuration being much more pronounced than in its in-plane counterpart (and the peak of the low-energy Maxwellian being correspondingly cooler). However, the out-of-plane spectrum for $\theta=30\deg$ is in good agreement with a 3D simulation having 64 cells along each transverse dimension (green line in panel \tit{b} of Fig.~\fig{specphi}), which suggests that results from our 2.5D out-of-plane simulations may be confidently applied to realistic three-dimensional scenarios. 

Concerning the fact that the $\theta=30\deg$ spectrum at $\ompt=9000$ shows a significant difference at high energies between in-plane and out-of-plane configurations,  we report that the spectrum computed in the shock transition layer (not shown) is remarkably similar between the two cases, suggesting that it is not the intrinsic acceleration efficiency to differ but rather the transmission probability of shock-accelerated particles into the downstream medium.

\section{B) The sharp transition in acceleration efficiency between subluminal and superluminal shocks}
In Fig.~\fig{spectot} we have shown that only subluminal shocks present a pronounced suprathermal tail in downstream particle spectra, whereas the particle distribution function behind superluminal shocks does not significantly deviate from a Maxwellian. Here, we show that the drop in acceleration efficiency between subluminal and superluminal configurations happens remarkably close to the boundary $\thetacrit\approx34\deg$ predicted by MHD calculations, and it is rather abrupt. 

\begin{figure*}[htbp]
\begin{center}
\includegraphics[width=0.75\textwidth]{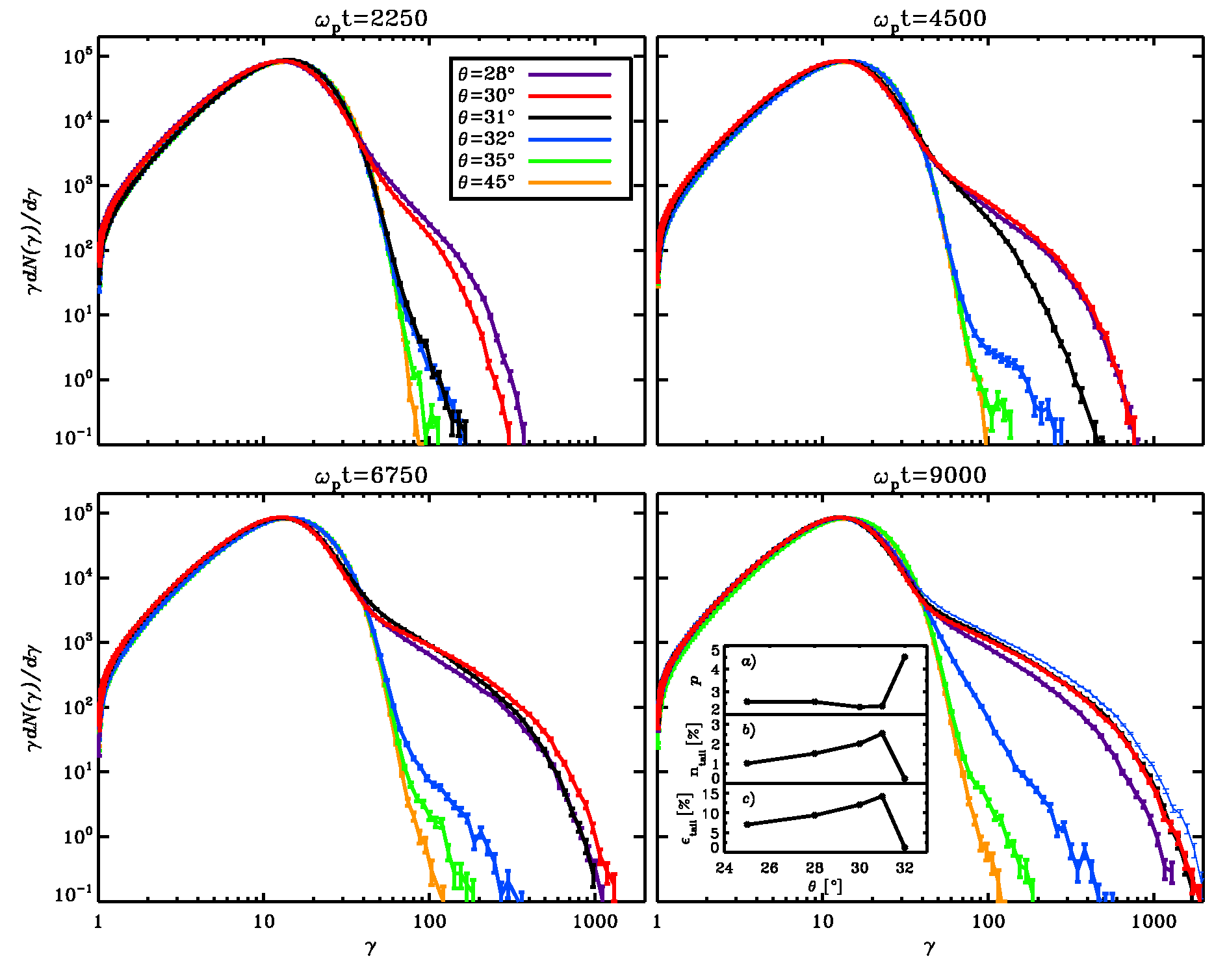}
\caption{Time evolution of downstream particle spectra for different obliquities, across the boundary $\theta_{\rm crit}\approx34^\circ$ between subluminal and superluminal shocks: $\theta=28^\circ$ (violet), $30^\circ$ (red), $31^\circ$ (black), $32^\circ$ (blue), $35\deg$ (green) and $45\deg$ (yellow). The subpanels \tit{a})-\tit{c}) in the last panel show at time $\ompt=9000$ the power-law slope of the suprathermal tail and the fraction of particles and energy stored in the tail, as a function of the obliquity angle $\theta$. The thin blue line in the bottom right panel shows the particle spectrum for $\theta=32\deg$ at $\ompt=13500$.}
\label{fig:specpart}
\end{center}
\end{figure*}

In Fig.~\fig{specpart}, downstream spectra for $\theta=28^\circ$ (violet), $30^\circ$ (red), $31^\circ$ (black), $32^\circ$ (blue), $35\deg$ (green) and $45\deg$ (yellow) are compared for different times. Within the subluminal range, lower obliquities ($\theta=28\deg$) show the most significant suprathermal tail  at early times, whereas shocks with magnetic inclinations closer to $\theta_{\rm crit}$ ($\theta=30\deg$ and $31\deg$) are the most efficient accelerators at later times. The closer to the superluminality threshold, the faster the time evolution of the spectrum, in agreement with the fact that particle acceleration in shocks with $\theta\lesssim\thetacrit$ is mostly mediated by SDA, which is a very fast process \citep{jokipii_82}. More importantly, the onset of efficient SDA is extremely rapid, since it is driven by a positive feedback, as discussed in \S\ref{sec:part30}: as the number of accelerated particles which escape upstream gets large enough to trigger the generation of the upstream oblique waves, a larger fraction of the incoming particles will be deflected from their straight path and their pitch angle at the shock will more likely satisfy the relation in \eq{inj}. More particles will then be reflected upstream and injected into the acceleration process, in a runaway cascade towards larger acceleration efficiency. The dramatic difference observed in the high-energy tail of $\theta=31\deg$ between $\ompt=2250$ and $\ompt=4500$ and of $\theta=32\deg$ between $\ompt=6750$ and $\ompt=13500$ (thin blue line in the bottom right panel of Fig.~\fig{specpart}) corresponds  to the onset of efficient SDA. In response to the increased acceleration efficiency, the peak of the low-energy Maxwellian shifts to lower temperatures. 

Since the positive feedback required for the onset of efficient SDA relies on upstream waves triggered by the returning particles, we expect that in superluminal shocks, where particles are not able to return upstream along the magnetic field, acceleration should be strongly suppressed. Indeed, the downstream spectra of the superluminal shocks $\theta=35^\circ$ and $\theta=45^\circ$ are mostly thermal, although a minor high-energy tail, further suppressed as $\theta$ increases, is observed at late times. Since spectra computed ahead of the shock (not plotted) do not show any sign of high-energy particles, we believe that the residual suprathermal tail observed downstream is probably populated by particles shock-drift accelerated during a single shock crossing from upstream to downstream and then advected downstream. We are confident that these conclusions hold true  for times longer than the timespan of our simulations, for the following reasons: \textit{i}) between $\ompt=6750$  and $\ompt=9000$, the downstream spectra of superluminal shocks do not significantly evolve (see also Fig.~\fig{spectime45}), so that they are presumably close to a steady-state; \textit{ii}) in the whole timespan explored by our simulations, there is no evidence for energetic particles escaping ahead of superluminal shocks, so that no upstream waves should be generated and the onset of efficient SDA should never occur.

The subpanels in the last panel of Fig.~\fig{specpart} show at time $\ompt=9000$ the power-law slope of the suprathermal tail and the fraction of particles and energy stored in the tail. We see that, with increasing magnetic obliquity within the subluminal range, the high-energy tail becomes flatter, more populated and more energetic. We remark that $\theta=32\deg$, which is still subluminal, does not accelerate very efficiently yet at $\ompt=9000$, but it will finally show a suprathermal tail similar to $\theta=30\deg$ and $\theta=31\deg$ (see the thin blue line  in the bottom right panel of Fig.~\fig{specpart}, for $\ompt=13500$). We also report that the downstream spectra for $\theta=30\deg$ and $\theta=31\deg$ do not significantly evolve between $\ompt=9000$ and $\ompt=13500$ (not shown), apart from the linear increase of the high-energy cutoff. This suggests that the timespan of our simulations can reasonably capture the evolution of  $\theta\lesssim\thetacrit$ shocks until they reach a steady-state, and that the values we quote in the subpanels of Fig.~\fig{specpart} (and Fig.~\fig{spectot} as well) are close to the saturation values.

\end{document}